\documentclass[aps,prb,showpacs,reprint,floatfix]{revtex4-1}
\pdfoutput=1  
\usepackage{graphicx,psfrag,epsf,epsfig}
\usepackage{dcolumn}
\usepackage{bm,bbm}
\usepackage{pstricks,pst-plot}     
\usepackage{latexsym}
\usepackage{mathrsfs,amsmath,amssymb,textcomp}
\usepackage{tikz}
\usetikzlibrary{matrix,arrows} 
\usepackage{hyperref}
\usepackage{float}

\def\d{{\mathrm d}}

\newcommand{\pf}[1]{\frac{\partial}{\partial #1}}

\begin{document}

\title{Quantum-kinetic theory of photocurrent generation via direct and 
phonon-mediated optical transitions}

\author{U. Aeberhard}
\email{u.aeberhard@fz-juelich.de}

\affiliation{IEF-5: Photovoltaik, Forschungszentrum J\"ulich, D-52425 J\"ulich,
Germany}

\date{\today}

\begin{abstract}
 A quantum kinetic theory of direct and phonon mediated indirect optical transitions is
 developed within the framework of the non-equilibrium Green's function formalism. After
 validation against the standard Fermi-Golden-Rule approach in the bulk case,
 it is used in the simulation of photocurrent generation in ultra-thin crystalline silicon 
 $p$-$i$-$n$-junction devices.
\end{abstract}


\maketitle

\section{\label{sec:intro}Introduction} 
 Quantum effects in semiconductor nanostructures are widely exploited in
 optoelectronic devices such as light emitting diodes or lasers. With the
 increasing demand of renewable energy supply, high efficiency photovoltaic
 devices were proposed, which also make use of semiconductor nanostructures \cite{green:00,green:01}. 
 However, these devices are not based on the standard III-V direct
 band gap materials conventionally used in optoelectronics, but on silicon, which is most common
 in solar cell applications. It was shown that even though the
 suppression of direct optical transitions is somewhat relaxed in low
 dimensional silicon structures, it is the phonon assisted indirect processes
 that dominate the optical response \cite{delerue:01}, and thus a suitable
 theoretical description of the photovoltaic behaviour needs to include these processes
 responsible for energy and momentum transfer, in addition to the general
 requirements of a quantum photovoltaic device, which were addressed in
[\onlinecite{ae:prb_08}]. As an extension of this previous work, the present
 paper presents a microscopic description of phonon-mediated optical transitions in 
 indirect semiconductor nanostructures, based on the non-equilibrium Green's function (NEGF) formalism
 and thus compatible with advanced quantum transport theories.
 
 The paper is organized as follows.  After the derivation of the general
 expressions for optoelectronic rates  within the non-equilibrium
 Green's function theory, these are applied to the case of a direct gap bulk
 semiconductor and compared to the standard Fermi-Golden-Rule (FGR) result.
 Next, the description of phonon-mediated indirect transitions within the NEGF formalism
 is discussed and the resulting expression for the generation rate again
 compared to the FGR, both in an analytical approximation and a numerical
 computation for a simple effective mass model of bulk silicon. Finally, the
 theory is extended to thin films in order to model photocurrent generation in
 ultrathin indirect gap semiconductor $p$-$i$-$n$ junctions, which form important building blocks of 
 future nano-photovoltaic devices.

\section{\label{sec:model}NEGF Theory}
In order to be able to describe both optical transitions and inelastic
quantum transport in semiconductor nanostructures within the same microscopic
picture, a theoretical framework based on the NEGF formalism was developed in 
[\onlinecite{ae:prb_08}]. Before considering the specific case of photogeneration in indirect semiconductors,
the aspects of the theory related to interband transitions will be outlined below in a 
general formulation, which relies on the extensive work on similar systems 
investigated with focus on their light-emitting and lasing properties
\cite{henneberger:88_3,jahnke:95,henneberger:96,pereira:96,pereira:98,henneberger:09pra,
steiger:thesis,steiger:iwce_09}.

\subsection{Interband generation and recombination rates}
The macroscopic equation of motion for a photovoltaic system is
the continuity equation for the charge carrier density
\begin{align}\partial_{t}\rho_{c}(\mathbf{r})+
\nabla\cdot
\mathbf{j}_{c}(\mathbf{r})=\mathcal{G}_{c}(\mathbf{r})-\mathcal{R}_{c}(\mathbf{r}),\quad
c=e,h
\label{eq:macro_eqmot} 
\end{align}
where $\rho_{c}$ and $j_{c}$ are charge carrier and particle current density,
respectively, $\mathcal{G}_{c}$ the generation rate and $\mathcal{R}_{c}$ the
recombination rate of carriers species $c$\footnote{The dimensions are that of
a volume rate, $[\mathcal{G},\mathcal{R}]=m^{-3}s^{-1}$.}.
The microscopic
version of this equation in terms of NEGF corresponds to \cite{kadanoff:62,keldysh:65}
($\underbar{1}\equiv\{\mathbf{r}_{1},t_{1}\in\mathcal{C}\}$)
\begin{align}
&\lim_{2\rightarrow 1}\Big\{i\hbar
\left(\pf{t_{1}}+\pf{t_{2}}\right)G(\underbar{1},\underbar{2})+\left[H_{0}(\mathbf{r_{1}})-
H_{0}(\mathbf{r_{2}})\right]G(\underbar{1},\underbar{2})\Big\}\nonumber\\
=&\lim_{2\rightarrow 1}\int_{\mathcal{C}}\d
3\left[\Sigma(\underbar{1},\underbar{3})G(\underbar{3},\underbar{2})-G(\underbar{1},\underbar{3})
\Sigma(\underbar{3},\underbar{2})\right]
,\label{eq:micro_eqmot}
\end{align}
where $H_{0}$ is the Hamiltonian of the non-interacting electronic system, $G$
is the electronic non-equilibrium Green's function defined on the
Keldysh contour $ \mathcal{C}$ and $\Sigma$ is the self-energy encoding the
interaction of the electronic system with phonons, photons and with itself, i.e.
the scattering processes which may give rise to intra- or interband transitions.
In steady state, \eqref{eq:macro_eqmot} becomes
\begin{align}
\nabla\cdot
\mathbf{j}_{c}(\mathbf{r})=\mathcal{G}_{c}(\mathbf{r})-\mathcal{R}_{c}(\mathbf{r}),\quad
c=e,h.
\label{eq:macro_cont}
\end{align}
In the microscopic theory, the divergence of the electron (particle) current
corresponds to the limit $1\rightarrow 2$ of the RHS in
\eqref{eq:micro_eqmot}\footnote{The factor of two is due to the assumption
of spin degeneracy.},
\begin{align}
\nabla\cdot 
\mathbf{j}(\mathbf{r})
=&-\frac{2}{V}\int\frac{dE}{2\pi\hbar}\int d^3 r'\Big[\Sigma^{R}
(\mathbf{r},\mathbf{r}';E)G^{<}(\mathbf{r}',\mathbf{r};E)\nonumber\\
&+\Sigma^{<}(\mathbf{r},\mathbf{r}';E)G^{A}(\mathbf{r}',\mathbf{r};E)
-G^{R}(\mathbf{r},\mathbf{r}';E)\nonumber\\&\times\Sigma^{<}(\mathbf{r}',\mathbf{r};E)
-G^{<}(\mathbf{r},\mathbf{r}';E)\Sigma^{A}(\mathbf{r}',\mathbf{r};E)\Big].\label{eq:currcons}
\end{align}
If the energy integration is taken over the
range of all bands connected by the interband scattering process, i.e. both
valence and conduction bands, the resulting divergence should vanish, if the
self-energies are chosen properly, ensuring the conservation of the overall
current, which is nothing else than the corresponding formulation of the
detailed balance requirement. If the integration is restricted to one of the bands 
(e.g. B=valence band or conduction band), the above equation corresponds to the microscopic version of 
\eqref{eq:macro_cont},
 and provides on the RHS  the total \emph{local} interband
scattering rate,
\begin{align}
R_{rad}(\mathbf{r})\equiv&\mathcal{R}(\mathbf{r})-\mathcal{G}(\mathbf{r})\\
=&-\frac{2}{V}\int_{B(\mathbf{r})}\frac{dE}{2\pi\hbar}\int
d^3 r'\Big[\Sigma^{R} (\mathbf{r},\mathbf{r}';E)G^{<}(\mathbf{r}',\mathbf{r};E)
\nonumber\\&+\Sigma^{<}(\mathbf{r},\mathbf{r}';E)G^{A}(\mathbf{r}',\mathbf{r};E)
-G^{R}(\mathbf{r},\mathbf{r}';E)\nonumber\\
&\times\Sigma^{<}(\mathbf{r}',\mathbf{r};E)-G^{<}(\mathbf{r},\mathbf{r}';E)\Sigma^{A}(\mathbf{r}',\mathbf{r};E)\Big]
\end{align}
Depending on the nature of the interaction described
by $\Sigma$, the scattering process may be highly non-local, in which case the
above rate contains contributions from a large volume. 
The total interband current is found by integrating the
divergence over the active volume, and is equivalent to the total
\emph{global} transition rate and, via the Gauss theorem, to the
difference of the interband currents at the boundaries of the interacting
region. Making use of the cyclic property of the trace, it can be
expressed in the form
\begin{align}
R_{rad}=&\frac{2}{V}\int d^3 r\int_{B(\mathbf{r})}\frac{dE}{2\pi\hbar}\int d^3
r' \Big[\Sigma^{<}(\mathbf{r},\mathbf{r}';E)G^{>}(\mathbf{r}',\mathbf{r};E)\nonumber\\
&-\Sigma^{>}(\mathbf{r},\mathbf{r}';E)G^{<}(\mathbf{r}',\mathbf{r};E)
\Big],\label{eq:totrate}
\end{align}
with units $[R_{rad}]=s^{-1}$.
Within the general (basis-independent) NEGF picture, The RHS of Eq. \eqref{eq:totrate} can be interpreted as follows: 
$\hbar^{-1}\Sigma^{<(>)}(E)$ represents the rate at which charge carriers with energy $E$ may leave (occupy)  a state at that energy
whereas $G^{>(<)}(E)$ quantifies the energy resolved probability that
the system can accept (donate) a particle of energy $E$, i.e. that there is an
unoccupied (occupied) state at the right energy. The first term thus represents
the total \emph{inscattering} rate, while the second term provides the total
\emph{outscattering} rate. 

If we are interested in the interband scattering rate, we can
neglect in Eq. \eqref{eq:totrate} the contributions to the
self-energy from \emph{intraband} scattering, e.g. via interaction with phonons,
low energy photons (free carrier absorption) or ionized impurities, since
they cancel upon energy integration over the band. However, if self-energies
and Green's functions are determined self-consistently as they should in
order to guarantee current conservation, the Green's functions are related to
the scattering self-energies via the Dyson equation for the propagator,

\begin{align}
 G^{R(A)}({\mathbf r_{1}},{\mathbf
 r}_{1'};E)=&G_{0}^{R(A)}({\mathbf r_{1}},{\mathbf
 r}_{1'};E)\nonumber\\& +\int d^{3}r_{2}\int
 d^{3}r_{3}G_{0}^{R(A)}({\mathbf r}_{1},
 {\mathbf r}_{2};E)\nonumber\\&\times\Sigma^{R(A)}({\mathbf r}_{2},{\mathbf
 r}_{3};E) G^{R(A)}({\mathbf r}_{3},{\mathbf
 r}_{1'};E),\label{eq:dyson1}
 \end{align}
 
 and the Keldysh equation for the correlation functions,
 \begin{align}
 G^{\lessgtr}({\mathbf r}_{1},{\mathbf
 r}_{1'};E)=&\int d^{3}r_{2}\int d^{3}r_{3} 
 G^{R}({\mathbf
 r}_{1}, {\mathbf r}_{2};E)\nonumber\\
 &\times \Sigma^{\lessgtr}({\mathbf
 r}_{2},{\mathbf r}_{3};E) G^{A}({\mathbf
 r}_{3},{\mathbf r}_{1'};E),\label{eq:dyson2}
\end{align}
and will thus be modified due to the intraband scattering. This means that in
the case of self-consistent solutions, it is in general not possible to
completely separate the effects of the different scattering processes, nor to isolate coherent from incoherent transport. 

In the remainder of the paper, the general theory of interband
transitions outlined above will be applied to optical interband transitions in
direct and indirect semiconductors. For computational purposes and to ease comparison with existing descriptions,
the theory will be reformulated using a simple effective mass band basis for completely homogeneous bulk systems and
for inhomogeneous thin film devices. 

\subsection{Bulk semiconductor}

 For a homogeneous bulk system, the field operators for carriers in
band $b$ can be written in the Bloch state basis,
\begin{align}
\hat{\Psi}_{b}(\mathbf{r},t)=\sum_{\mathbf{k}}\psi_{b\mathbf{k}}(\mathbf{r})\hat{c}_{b\mathbf{k}}(t).\label{eq:bulkfieldop}
\end{align}
The expression for the total radiative rate of carriers in
band $b$ is simplified by using the Fourier space representation of Green's
functions and self-energies, which reads ($O=G,\Sigma$)
\begin{align}
O_{b,b'}(\mathbf{r},\mathbf{r'};E)=\sum_{\mathbf{k}}\psi_{b\mathbf{k}}(\mathbf{r})O_{b,b'}(\mathbf{k};E)\psi_{b'\mathbf{k}}^{*}(\mathbf{r'}),
\end{align}
where $O_{b,b'}(\mathbf{k};E)$ is the steady-state Fourier-transform of the real-time Keldysh components of the contour-ordered Green's function
\begin{align}
\mathcal{O}_{b,b'}(\mathbf{k};t-t')=\frac{1}{i\hbar}\left\langle
\hat{T}_{\mathcal{C}}\left\{\hat{c}_{b\mathbf{k}}(t) \hat{c}_{b'\mathbf{k}}^{\dagger}(t')\right\}\right\rangle.
\end{align}
Inserting these expressions in \eqref{eq:totrate}, the band-resolved rates are
obtained as
\begin{align}
R_{rad,b}=\frac{2}{V}\int\frac{dE}{2\pi\hbar}\sum_{\mathbf{k}}\sum_{b'}
\Big[&\Sigma_{b,b'}^{<}(\mathbf{k};E)G_{b',b}^{>}(\mathbf{k};E)\nonumber\\&
-\Sigma_{b,b'}^{>}(\mathbf{k};E)G_{b',b}^{<}(\mathbf{k};E)\Big].
\end{align}
In the following, the off-diagonal terms will be neglected ($O_{b}\equiv O_{b,b'}$), which means that only incoherent interband and subband
polarization is considered. 

\subsubsection{\label{sec:dirintrans}Direct interband transitions}
Inserting the electron-photon self-energy for a
two-band model of a direct semiconductor (App. \ref{sec:appA}), and neglecting
intraband processes (free-carrier absorption and emission), the expression for the interband absorption 
rate becomes (in the following, the energy integration is restricted to the conduction band) 
\begin{align}
R_{abs}=&\frac{2}{V}\int\frac{dE}{2\pi\hbar}\sum_{\mathbf{k}}
\sum_{\lambda,\mathbf{q}}|M^{\gamma}_{cv}(\mathbf{k},\lambda,\mathbf{q})|^{2}
N^{\gamma}_{\lambda,\mathbf{q}}\nonumber\\&\times G_{v}^{<}(\mathbf{k};E-\hbar\omega_{\mathbf{q}})G_{c}^{>}
(\mathbf{k};E),\label{eq:directopttrans}
\end{align}
with $M^{\gamma}_{cv}$ the optical matrix element (App. \ref{sec:appA}) and $N^{\gamma}_{\lambda,\mathbf{q}}$  
the occupation of the photon modes. The latter is obtained from the modal photon flux via
$N^{\gamma}_{\lambda,\mathbf{q}}=\phi^{\gamma}_{\lambda,\mathbf{q}}V/\tilde{c}$,
where $\tilde{c}$ is the speed of the light in the active medium. The modal photon flux in turn is given by the 
modal intensity of the EM field  as $\phi^{\gamma}_{\lambda,\mathbf{q}}=I^{\gamma}_{\lambda,\mathbf{q}}/(\hbar\omega_{\lambda,\mathbf{q}})$.
Similarly, stimulated interband emission reads 
\begin{align}
R_{em,st}=&\frac{2}{V}\int\frac{dE}{2\pi\hbar}\sum_{\mathbf{k}}
\sum_{\lambda,\mathbf{q}}|M^{\gamma}_{cv}(\mathbf{k},\lambda,\mathbf{q})|^{2}N^{\gamma}_{\lambda,\mathbf{q}}\nonumber\\&\times
G_{v}^{>}(\mathbf{k};E-\hbar\omega_{\mathbf{q}})G_{c}^{<}(\mathbf{k};E),
\end{align}
while spontaneous interband emission is expressed as 
\begin{align}
R_{em,sp}=&\frac{2}{V}\int\frac{dE}{2\pi\hbar}\sum_{\mathbf{k}}
\sum_{\lambda,\mathbf{q}}|M^{\gamma}_{cv}(\mathbf{k},\lambda,\mathbf{q})|^{2}\nonumber\\&\times
G_{v}^{>}(\mathbf{k};E-\hbar\omega_{\mathbf{q}})G_{c}^{<}(\mathbf{k};E).
\end{align}
The net absorption $R_{abs,net}=R_{abs}-R_{em,st}$ of photons in mode
$(\lambda,\mathbf{q})$ is thus given by
\begin{align}
R_{abs,net}(\lambda,\mathbf{q})=&\frac{2}{V}\int\frac{dE}{2\pi\hbar}\sum_{\mathbf{k}}
|M^{\gamma}_{cv}(\mathbf{k},\lambda,\mathbf{q})|^{2}N^{\gamma}_{\lambda,\mathbf{q}}\nonumber\\
&\times\Big[G_{v}^{<}(\mathbf{k};E-\hbar\omega_{\mathbf{q}})G_{c}^{>}(\mathbf{k};E)\nonumber\\
&~~~~-G_{v}^{>}(\mathbf{k};E-\hbar\omega_{\mathbf{q}})G_{c}^{<}(\mathbf{k};E)\Big].\label{eq:gfrate}
\end{align}

Conventionally, the rates for absorption and emission are calculated based on
Fermi's Golden Rule (FGR), corresponding to the Born approximation within
first order perturbation theory. It is thus instructive to compare the above
expressions with the FGR-rate for net direct interband absorption \cite{bassani:75},
\begin{align}
R^{FGR}_{abs,net}(\lambda,\mathbf{q})=&\frac{2}{V}\sum_{\mathbf{k}}\frac{2\pi}{\hbar}
|M^{\gamma}(\mathbf{k},\lambda,\mathbf{q})|^{2}
N^{\gamma}_{\lambda,\mathbf{q}}\nonumber\\&\times\delta\big(\varepsilon_{c}(\mathbf{k})
-\varepsilon_{v}(\mathbf{k})-\hbar\omega_{\mathbf{q}}\big)
\big[f_{v}(\mathbf{k})-f_{c}(\mathbf{k})\big].\label{eq:fgrrate}
\end{align}
Here, the absorbing (bulk) material is described by the dispersion relations $\varepsilon_{b}(\mathbf{k})$, $b=c,v$, and assumed to be in a
quasi-equilibrium state with occupation described by the Fermi function $f_{b}(E)=(\exp[\beta(E-\mu_{b})]+1)^{-1}$,
$\beta=(k_{B}T)^{-1}$, where $\mu_{b}$, $b=c,v$ are global quasi-Fermi levels.

In order to reproduce the FGR result from the more general expression in terms of Green's functions, the un-perturbed (i.e. non-interacting) 
equilibrium form of the latter needs to be used, which corresponds to the expressions for free fermions in equilibrium,  
\begin{align}
G_{b}^{<(0)}(\mathbf{k};E)=&2\pi i
f_{b}(E)\delta\big(E-\varepsilon_{b}(\mathbf{k})\big),\label{eq:equil_gf_1}\\
G_{b}^{>(0)}(\mathbf{k};E)=&2\pi i
\big[f_{b}(E)-1\big]\delta\big(E-\varepsilon_{b}(\mathbf{k})\big),
\label{eq:equil_gf_2}
\end{align} 
Introducing these expressions in \eqref{eq:gfrate} and carrying-out the energy
integration reproduces the FGR expression \eqref{eq:fgrrate}. 

To estimate the deviation of the rate from the FGR result for self-energies
beyond the first Born approximation, the FGR rate is first used to derive the
standard expression for the bulk absorption coefficient of the two band
effective mass model, which amounts to 
\begin{align}
\alpha(\hbar\omega_{\gamma})=\frac{R_{abs,net}(\hbar\omega_{\gamma})/V}{S(\hbar\omega_{\gamma})/\hbar\omega_{\gamma}},
\end{align}
where $S(\hbar\omega_{\gamma})$ is the monochromatic energy flux density of the
EM field (i.e. the absolute value of the Poynting vector) given by 
\begin{align}
S(\hbar\omega_{\gamma})=\rho_{\gamma}(\hbar\omega_{\gamma})\hbar\omega_{\gamma}\varepsilon_{b}\tilde{c}\int\frac{d\Omega}{4\pi}
\sum_{\lambda}N^{\gamma}_{\lambda}(\hbar\omega_{\gamma},\Omega),
\end{align}
where
\begin{align}
\rho_{\gamma}(\hbar\omega_{\gamma})=\frac{(\hbar\omega_{\gamma})^2}{2\pi^2(\hbar\tilde{c})^3}
\end{align}
is the photonic density of states of an optically isotropic medium with
refractive index $n_{b}=\sqrt{\varepsilon_{b}}$ and corresponding speed of light
$\tilde{c}=c_{0}/n$.  With the standard approximation of isotropic and
momentum-independent optical matrix elements, i.e.
$|M^{\gamma}_{cv}(\mathbf{k},\lambda,\mathbf{q})|^2\approx\bar{\mathcal{M}}^{\gamma}_{cv}(\hbar\omega_{\gamma})$,
the absorption rate is rewritten as follows,
\begin{align}
R_{abs}(\hbar\omega_{\gamma})=&\bar{\mathcal{M}}^{\gamma}_{cv}(\hbar\omega_{\gamma})
\mathcal{J}_{cv}(\hbar\omega_{\gamma})\rho_{\gamma}(\hbar\omega_{\gamma})
\nonumber\\&\times
\int\frac{d\Omega}{4\pi}\sum_{\lambda}N^{\gamma}_{\lambda}(\hbar\omega_{\gamma},\Omega)\\ 
=&\bar{\mathcal{M}}^{\gamma}_{cv}(\hbar\omega_{\gamma})\mathcal{J}_{cv}(\hbar\omega_{\gamma})
S(\hbar\omega_{\gamma})/(\hbar\omega_{\gamma}\tilde{c}),
\end{align}
which provides the bulk absorption coefficient 
\begin{align}
\alpha(\hbar\omega_{\gamma})=&\tilde{\mathcal{M}}^{\gamma}_{cv}(\hbar\omega_{\gamma})
\mathcal{J}_{cv}(\hbar\omega_{\gamma}),
\end{align}
with
$\tilde{\mathcal{M}}^{\gamma}_{cv}=\bar{\mathcal{M}}^{\gamma}_{cv}/\tilde{c}$.
The difference between FGR and NEGF approaches concerns the term
$\mathcal{J}_{cv}(\hbar\omega_{\gamma})$, which in the FGR case takes the
specific form (using the continuum approximation $\sum_{\mathbf{k}}\rightarrow \frac{V}{(2\pi)^3}\int
d^3 k$ )
\begin{align}
\mathcal{J}_{cv}^{FGR}(\hbar\omega_{\gamma})=&\frac{2\pi}{\hbar}\frac{2}{(2\pi)^3}\int
d^3 k~\delta\big(\varepsilon_{c}(\mathbf{k})
-\varepsilon_{v}(\mathbf{k})-\hbar\omega_{\gamma}\big)\nonumber\\&\times
\big[f_{v}(\mathbf{k})-f_{c}(\mathbf{k})\big].\label{eq:jdos_fgr}
\end{align}
If the occupation depends only marginally on crystalline momentum, 
the above expressions related to the joint density
of states $J_{cv}$ with suitable occupation,
\begin{align}
\mathcal{J}_{cv}^{FGR}(\hbar\omega_{\gamma})=&\frac{2\pi}{\hbar}J_{cv}(\hbar\omega_{\gamma})
\big(f_{v}-f_{c}\big),\\
J_{cv}(\hbar\omega_{\gamma})=&\frac{2}{(2\pi)^3}\int
d^3 k~\delta\big(\varepsilon_{c}(\mathbf{k})
-\varepsilon_{v}(\mathbf{k})-\hbar\omega_{\gamma}\big).\label{eq:jdos_fgr_dir)}
\end{align}
In the NEGF case, joint density of states and occupation cannot be separated, but are both
contained in the Keldysh Green's functions
\begin{align}
\mathcal{J}_{cv}^{GF}(\hbar\omega_{\gamma})=&\frac{2}{(2\pi)^3}\int
d^3 k\int\frac{dE}{2\pi\hbar}
\Big[G_{v}^{<}(\mathbf{k};E-\hbar\omega_{\gamma})G_{c}^{>}(\mathbf{k};E)\nonumber\\
&-G_{v}^{>}(\mathbf{k};E-\hbar\omega_{\gamma})G_{c}^{<}(\mathbf{k};E)\Big]\label{eq:jdosgf}\\
&=\hat{\mathcal{P}}_{cv,0}(\mathbf{0},\hbar\omega_{\gamma}),
\end{align}
which via
$\hat{\mathcal{P}}_{cv,0}=\mathcal{P}_{cv,0}^{>}-\mathcal{P}_{cv,0}^{<}$ is
related to the Keldysh components of the free-carrier interband  polarization function
\begin{align}
\mathcal{P}_{cv,0}^{\lessgtr}(\mathbf{q},E)=&\frac{2}{(2\pi)^3}\int
d^3
k\int\frac{d \tilde{E}}{2\pi\hbar}G_{c}^{\lessgtr}(\mathbf{k};\tilde{E})\nonumber\\
&\times G_{v}^{\gtrless}(\mathbf{k}-\mathbf{q};\tilde{E}-E).\label{eq:polfun}
\end{align}
This expression is valid in \emph{any} situation that can be described in
terms of single-particle Green's functions. The deviations from the FGR result are marginal in 
the quasi-equilibrium case and in absence of further interactions beyond electron-light coupling and of sources 
of non-equilibrium, which both modify the Green's functions, causing them to differ from
the expressions given in Eqs. \eqref{eq:equil_gf_1} and \eqref{eq:equil_gf_2}. It is straightforward to show that
inserting the latter expressions in \eqref{eq:jdosgf} reproduces
\eqref{eq:jdos_fgr}, and for the special case of spherical bands, the
joint density of states is given by the well-known analytical expression

 \begin{align} 
 J_{cv}^{dir}(\hbar\omega_{\gamma})=\frac{(2m_{r}^{*})^{\frac{3}{2}}}{\pi^2\hbar^3}\sqrt{E_{\gamma}-E_{g}},
\end{align}
where $m_{r}^{*}$ is the reduced effective mass and $E_{g}$ is the direct band
gap.

\subsubsection{Phonon assisted interband transitions}
\begin{figure}[h]
\begin{center}
\includegraphics[width=8cm]{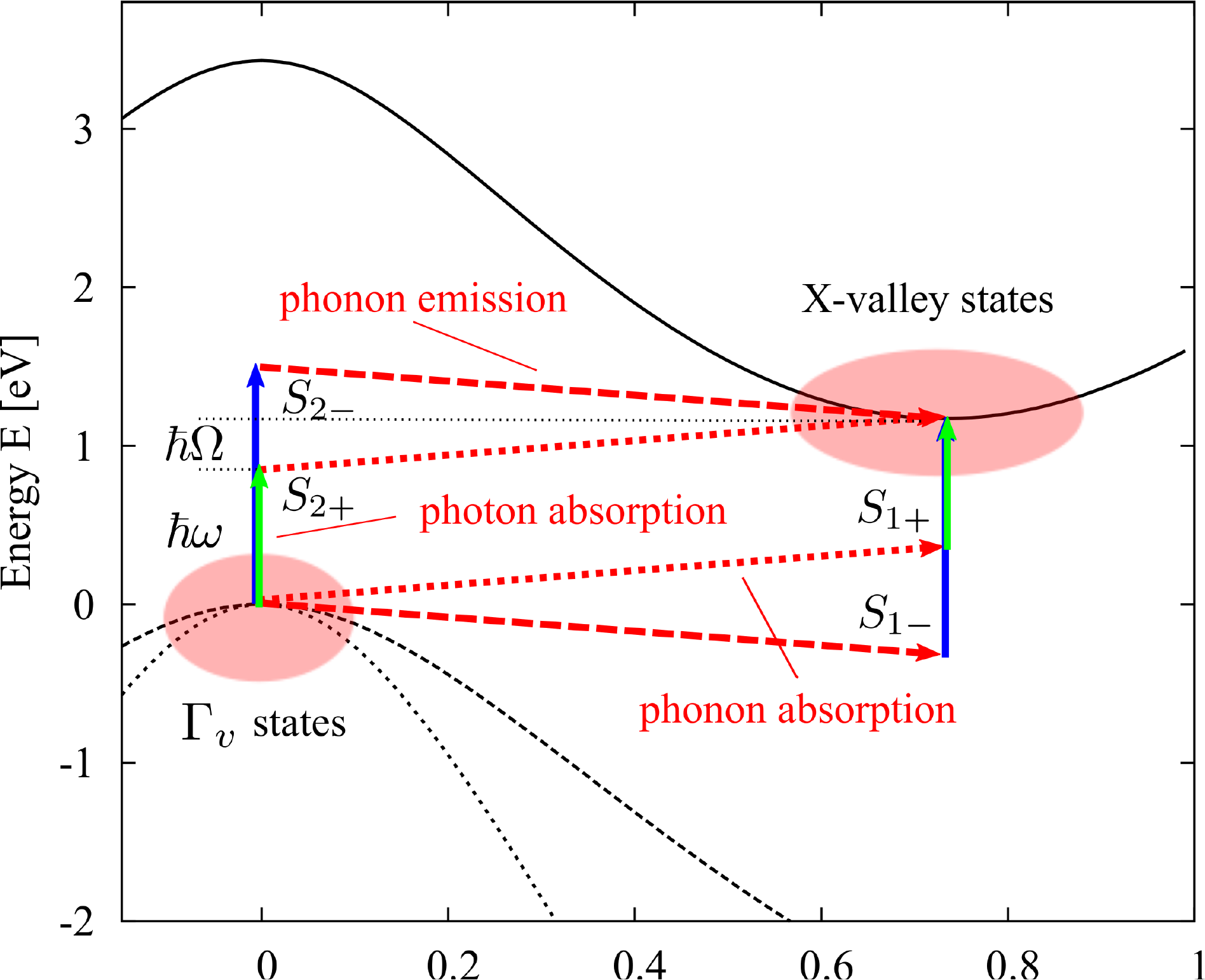}
\caption{Possible excitation pathways for optical interband transitions in indirect semiconductors on the example of bulk
silicon.\label{fig:indtrans}}
\end{center} 
\end{figure}

Even in the case of a direct semiconductor discussed above, the presence of phonons can have a considerable effect on the 
optical transition rates as new excitation paths become available, i.e. the phonons strongly increase the number of initial-final state pairs
separated by a given transition energy. However, comparing to the direct transition, this enhancement of the joint density of states is
overcompensated by the fact that the transitions become much more unlikely due to the need for coupling to a suitable phonon and the detour via the
virtual state. As a consequence, phonon assisted transitions may be neglected in direct bulk semiconductors where crystalline momentum is conserved.
Obviously, the situation is completely different in indirect bulk semiconductors, where no direct transitions are possible. There, at lowest
non-vanishing order, four different excitation processes exist (Fig. \ref{fig:indtrans}):
\begin{center}
\begin{tabular}{lll}
 $S_{1+}:$& 1. phonon absorption, &2. photon absorption, \\  
 $S_{1-}:$& 1. phonon emission, &2. photon absorption, \\  
 $S_{2+}:$& 1. photon absorption, &2. phonon absorption, \\  
 $S_{2-}:$& 1. photon absorption, &2. phonon emission.  
\end{tabular}
\end{center}

To exemplify the inclusion of the phonon scattering in the NEGF description of optical interband transitions, 
we will in the following focus on processes $S_{2\pm}$.
The FGR transition rate for these processes are obtained from second order
perturbation theory as
\begin{align}
R_{abs,\pm}^{FGR}(\lambda,\mathbf{q})=&\frac{2\pi}{\hbar}N^{\gamma}_{\lambda,\mathbf{q}}\sum_{\mathbf{k}_{c},\mathbf{k}_{v}}
\sum_{\Lambda,\mathbf{Q}}\frac{[M^{ph}_{\Lambda,\mathbf{Q}}(\mathbf{k}_{v},\mathbf{k}_{c})]^{2}
[M^{\gamma}_{\lambda,\mathbf{q}}(\mathbf{k}_{v})]^{2}
}{|\varepsilon_{c}(\mathbf{k}_{v})-\varepsilon_{v}(\mathbf{k}_{v})
-\hbar\omega_{\mathbf{q}}|^{2}}\nonumber\\
\times&\left[N_{\Lambda,\mathbf{Q}}^{ph}+\frac{1}{2}\mp\frac{1}{2}\right]
f_{v}(\mathbf{k}_{v})[1-f_{c}(\mathbf{k}_{c})]
\nonumber\\ 
\times
&\delta(\varepsilon_{c}(\mathbf{k}_{c})
-\varepsilon_{v}(\mathbf{k}_{v})
-\hbar\omega_{\mathbf{q}}\mp\hbar\Omega_{\Lambda,\mathbf{Q}}).\label{eq:fgr_indopt}
\end{align}
Here, $M^{ph}_{\Lambda,\mathbf{Q}}$ encodes the matrix element for coupling of electrons to the phonon mode $(\Lambda,\mathbf{Q})$ with 
energy $\hbar\Omega_{\Lambda,\mathbf{Q}}$ and occupation given by the Bose-Einstein distribution $N_{\Lambda,\mathbf{Q}}^{ph}=
\left(e^{\beta\hbar\Omega_{\Lambda,\mathbf{Q}}}-1\right)^{-1},~\beta=(k_{B}T)^{-1}$ at lattice temperature $T$.
In analogy to the direct case, with the previously used approximations for the optical matrix elements and the additional assumptions
\begin{align}
|\varepsilon_{c}(\mathbf{k}_{v})-\varepsilon_{v}(\mathbf{k}_{v})
-\hbar\omega_{\mathbf{q}}|&\approx |E_{g0}-\hbar\omega_{\mathbf{q}}|,\\
M^{ph}_{\Lambda,\mathbf{Q}}(\mathbf{k}_{v},\mathbf{k}_{c})&\approx\mathcal{M}^{ph}_{\Lambda}
\delta(\mathbf{k}_{v}-\mathbf{k}_{c}+\mathbf{Q})
\end{align}
we can write the (phonon-assisted) absorption coefficient
\begin{align}
\alpha^{ind}(\hbar\omega_{\gamma})=&\frac{\tilde{\mathcal{M}}^{\gamma}_{cv}(\hbar\omega_{\gamma})^2}
{|E_{g0}-\hbar\omega_{\gamma}|^2}\sum_{\Lambda}\mathcal{M}^{ph}_{\Lambda}\nonumber\\
&\times\sum_{s=\pm}
\left[N_{\Lambda}^{ph}+\frac{1}{2}-s\frac{1}{2}\right]\mathcal{J}_{cv,s}^{ind}(\hbar\omega_{\gamma}),
\end{align} 
but where now the joint density of states for indirect transitions is used in
\eqref{eq:jdos_fgr},
\begin{align}
J_{cv,\pm}^{ind}(\hbar\omega_{\gamma})=&\frac{2}{(2\pi)^6}\int
d^3 k_{1}\int
d^3 k_{2}~\delta\big(\varepsilon_{c}(\mathbf{k}_{1})
-\varepsilon_{v}(\mathbf{k}_{2})\nonumber\\&
-\hbar\omega_{\gamma}\mp\hbar\Omega_{\Lambda,\mathbf{k}_{1}-\mathbf{k}_{2}}\big),\label{eq:jdos_fgr_ind}
\end{align}
which for spherical bands and a single phonon frequency $\Omega$ may be
simplified to
\begin{align}
J_{cv,\pm}^{ind}(\hbar\omega_{\gamma})=&\frac{(m_{c}^{*}m_{v}^{*})^{\frac{3}{2}}}{(2\pi
\hbar)^3}(\hbar\omega_{\gamma}-E_{g}\mp
\hbar\Omega)^2,\label{eq:jdos_fgr_ind_sper}
\end{align}
where $E_{g}$ denotes the (indirect) band gap.

In the following, the phonon-assisted absorption rate shall be derived within
the NEGF formalism, starting from the expression for the absorption rate, but where now the Green's functions 
contain the contributions due to the electron-phonon scattering. 
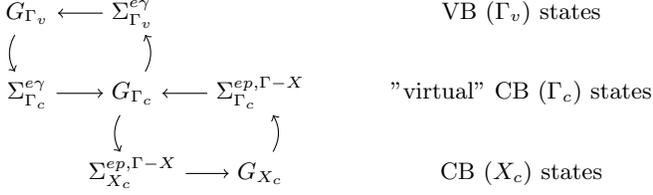
\begin{figure}[t]
 \begin{center}
 \begin{tikzpicture}
\matrix(m)[matrix of math nodes, row sep=5mm, column sep=3mm, text height=2.5mm,
text depth=1mm] {
G_{\Gamma_{v}}  \pgfmatrixnextcell \Sigma^{e\gamma}_{\Gamma_{v}}
\pgfmatrixnextcell ~ \pgfmatrixnextcell~\pgfmatrixnextcell\textnormal{VB
($\Gamma_{v}$) states} \\ \Sigma^{e\gamma}_{\Gamma_{c}}
\pgfmatrixnextcell
G_{\Gamma_{c}} \pgfmatrixnextcell \Sigma^{ep,\Gamma-X}_{\Gamma_{c}}
\pgfmatrixnextcell~\pgfmatrixnextcell\textnormal{''virtual'' CB ($\Gamma_{c}$)
states}\\ ~\pgfmatrixnextcell \Sigma^{ep,\Gamma-X}_{X_{c}}
\pgfmatrixnextcell G_{X_{c}}\pgfmatrixnextcell \pgfmatrixnextcell \textnormal{CB ($X_{c}$) states}\\};
\path[->]
(m-1-2) edge  [auto] (m-1-1)
(m-1-1) edge  [bend right] (m-2-1)
(m-2-2) edge  [bend right] (m-1-2)
(m-2-1) edge  [auto] (m-2-2)
(m-2-3) edge  [auto] (m-2-2)
(m-2-2) edge  [bend right] (m-3-2)
(m-3-2) edge  [auto] (m-3-3)
(m-3-3) edge  [bend right] (m-2-3);
\end{tikzpicture}
\caption{Self-consistent computation of Green's functions and scattering self-energies in silicon enabling the description of phonon-assisted
indirect optical transitions\label{fig:selfconsit}}
\end{center}
\end{figure}
Since the Green's functions and interaction self-energies are evaluated in a self-consistency iteration process, 
all possible single phonon processes are included to all orders, i.e. the Green's functions contain contributions 
from a number of scattering processes that corresponds to the number of self-consistency iteration steps. It is 
thus via this self-consistent computation that phonon assisted optical transitions are enabled, even though the 
self-energies themselves are only on the level of a first order self-consistent Born approximation, i.e. do not 
include the combination of both electron-phonon and electron-photon scattering at the same time. In Fig. 
\ref{fig:selfconsit}, this procedure is shown for the technologically relevant example of indirect interband 
absorption of photons in silicon, where zone-boundary phonons provide the wave-vector difference in a $\Gamma-X$ 
inter-valley scattering process. 

We start again from the general expression for the in-scattering rate, which in 
this case reads
\begin{align}
R_{in}=&\frac{2}{V}\int\frac{dE}{2\pi\hbar}\sum_{\mathbf{k}}
\Sigma_{X_{c}}^{<}(\mathbf{k};E)G_{X_{c}}^{>}(\mathbf{k};E).
\end{align}
Assuming a photon-first indirect process ($S_{2\pm}$), the corresponding
expression for $\Gamma-X$-scattering is inserted for the self-energy
(see Eq. \eqref{eq:phonintse} in App. \ref{sec:app_se}), leading to the
equivalent of Eq. \eqref{eq:directopttrans} ($\sigma=LA,TO$: phonon mode),
\begin{align}
R_{in}=&\frac{2}{V}\int\frac{dE}{2\pi\hbar}\sum_{\mathbf{k}}
\sum_{\mathbf{Q},\sigma}[M^{ph}_{\sigma}(\Omega_{\sigma})]^{2}\big[N^{ph}_{\sigma}
G_{\Gamma_{c}}^{<}(\mathbf{Q};E-\hbar\Omega_{\sigma})\nonumber\\&+(N^{ph}_{\sigma}+1)G_{\Gamma_{c}}^{<}(\mathbf{Q};E+\hbar\Omega_{\sigma})
G_{X_{c}}^{>}(\mathbf{k};E).
\end{align}
In the next step, the Keldysh equation for electron-photon interaction is used
to replace the lesser $\Gamma_{c}$-GF,
\begin{align}
G_{\Gamma_{c}}^{<}(\mathbf{k};E)=G_{\Gamma_{c}}^{R}(\mathbf{k};E)\Sigma_{\Gamma_{c}}^{<\gamma}(\mathbf{k};E)
G_{\Gamma_{c}}^{A}(\mathbf{k};E)
\end{align}
providing the modal absorption rate
\begin{align}
&R_{abs}(\lambda,\mathbf{q})=\frac{2}{V}\int\frac{dE}{2\pi\hbar}\sum_{\mathbf{k}}
\sum_{\mathbf{Q},\sigma}[M^{ph}_{\sigma}(\Omega_{\sigma})]^{2}[M^{\gamma}(\mathbf{k},\lambda,\mathbf{q})]^{2}\nonumber\\&\times N^{\gamma}_{\lambda,\mathbf{q}}
\Big[N^{ph}_{\sigma}G_{\Gamma_{c}}^{R}(\mathbf{Q};E-\hbar\Omega_{\sigma})
G_{\Gamma_{v}}^{<}(\mathbf{Q};E-\hbar\Omega_{\sigma}-\hbar\omega_{\mathbf{q}})\nonumber\\&\times G_{\Gamma_{c}}^{A}(\mathbf{Q};
E-\hbar\Omega_{\sigma})+(N^{ph}_{\sigma}+1)G_{\Gamma_{c}}^{R}(\mathbf{Q};E+\hbar\Omega_{\sigma})\nonumber\\
&\times
G_{\Gamma_{v}}^{<}(\mathbf{Q};E+\hbar\Omega_{\sigma}-\hbar\omega_{\mathbf{q}})G_{\Gamma_{c}}^{A}(\mathbf{Q};
E+\hbar\Omega_{\sigma})\Big]\nonumber\\
&\times G_{X_{c}}^{>}(\mathbf{k};E).\label{eq:indgfrate}
\end{align}
Again, this is to be compared to the FGR result, which, for the same
electron-photon and electron-phonon interaction Hamiltonian terms, follows from
second-order perturbation theory to
\begin{align}
R_{abs}^{FGR}(\lambda,\mathbf{q})=&\frac{2\pi}{\hbar}\sum_{\mathbf{k}_{c},\mathbf{k}_{v}}\sum_{\sigma}
\frac{[M^{ph}_{\sigma}(\Omega_{\sigma})]^{2}[M^{\gamma}(\mathbf{k}_{v},\lambda,\mathbf{q})]^{2}
N^{\gamma}_{\lambda,\mathbf{q}}}{|\varepsilon_{\Gamma_{c}}(\mathbf{k}_{v})-\varepsilon_{\Gamma_{v}}(\mathbf{k}_{v})
-\hbar\omega_{\mathbf{q}}|^{2}}\nonumber\\\times&\Big[N_{\sigma}^{ph}\delta(\varepsilon_{X_{c}}(\mathbf{k}_{c})
-\varepsilon_{\Gamma_{v}}(\mathbf{k}_{v})
-\hbar\omega_{\mathbf{q}}-\hbar\Omega_{\sigma})\nonumber\\+&(N_{\sigma}^{ph}+1)\delta(\varepsilon_{X_{c}}(\mathbf{k}_{c})
-\varepsilon_{\Gamma_{v}}(\mathbf{k}_{v})
-\hbar\omega_{\mathbf{q}}+\hbar\Omega_{\sigma})\Big]\nonumber\\\times
&f_{\Gamma_{v}}(\mathbf{k}_{v})[1-f_{X_{c}}(\mathbf{k}_{c})].\label{eq:fgr_indopt}
\end{align}
 Now, inserting the non-interacting equilibrium expressions for the lesser,
 greater and retarded GF in \eqref{eq:indgfrate} provides the expression
 \begin{widetext}
\begin{align}
R_{abs}(\lambda,\mathbf{q})=&\frac{2}{V}\int\frac{dE}{2\pi\hbar}\sum_{\mathbf{k}}
\sum_{\mathbf{Q},\sigma}[M^{ph}_{\sigma}(\Omega_{\sigma})]^{2}[M^{\gamma}(\mathbf{k},\lambda,\mathbf{q})]^{2}
N^{\gamma}_{\lambda,\mathbf{q}}\Bigg[\frac{N_{\sigma}^{ph}if_{\Gamma_{v}}(E-\hbar\Omega_{\sigma}-\hbar\omega_{\mathbf{q}})
2\pi\delta\big(E-\hbar\Omega_{\sigma}-\hbar\omega_{\mathbf{q}}-\varepsilon_{\Gamma_{v}}(\mathbf{Q})\big)}
{|E-\hbar\Omega_{\sigma}-\hbar\omega_{\mathbf{q}}-\varepsilon_{\Gamma_{c}}(\mathbf{Q})+i\eta_{0+}|^{2}}
\nonumber\\
+&\frac{(N_{\sigma}^{ph}+1)if_{\Gamma_{v}}(E+\hbar\Omega_{\sigma}-\hbar\omega_{\mathbf{q}})
2\pi\delta\big(E+\hbar\Omega_{\sigma}-\hbar\omega_{\mathbf{q}}-\varepsilon_{\Gamma_{v}}(\mathbf{Q})\big)}
{|E+\hbar\Omega_{\sigma}-\hbar\omega_{\mathbf{q}}-\varepsilon_{\Gamma_{c}}(\mathbf{Q})+i\eta_{0+}|^{2}}\Bigg]i[f_{X_{c}}(E)-1]
2\pi\delta\big(E-\varepsilon_{X_{c}}(\mathbf{k})\big),
\end{align}
\end{widetext}
 which, upon performing the energy integration and with
 $\mathbf{Q}=\mathbf{k}_{v}$, $\mathbf{k}=\mathbf{k}_{c}$, reproduces again the
 FGR result given in \eqref{eq:fgr_indopt}.   
 
 \begin{table}[b]
\caption{\label{tab:band_parameters} Material parameters used in simulations}
\begin{ruledtabular}
\begin{tabular}{cccccc}
 $m^{*}_{\Gamma c}$& $m^{*}_{X}$&$m^{*}_{\Gamma
 v}$&$E_{g,\Gamma v-\Gamma c}$&$E_{g,\Gamma v-X}$  &$P_{cv}^2/m_{0}$ \\\hline
 \\0.3 $m_{0}$& 0.3 $m_{0}$&0.54 $m_{0}$& 3.1 eV& 1.17 eV& 4 eV \\
 \\
 $\sigma$&Mode &$\hbar\Omega_{\sigma}$&$D_{iv}K_{\sigma}$&&\\
\noalign{\smallskip}\hline\noalign{\smallskip}
($\Gamma$-X$)_{1}$&LA& 18.4 meV&2.45$\times 10^8$ eV/cm &&\\
($\Gamma$-X$)_{2}$&TO& 57.6 meV&0.8 $\times 10^8$ eV/cm&&\\
\end{tabular}
\end{ruledtabular}
\end{table} 

\begin{figure}[b!]
\begin{center}
\includegraphics[width=7.5cm]{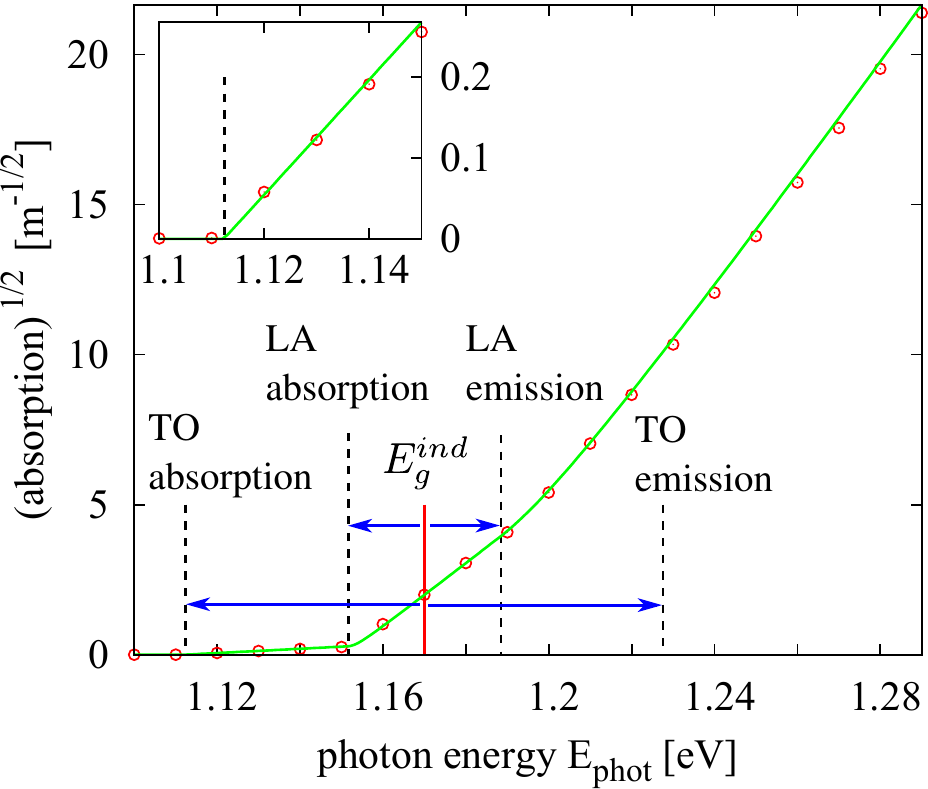}
\caption{Photon absorption rate for indirect optical transitions in intrinsic bulk silicon: FGR and NEGF rates are in close agreement, since the
correction fo the conduction band DOS due to the phonon-mediated $\Gamma-X$ scattering is weak.
\label{fig:indabs}}
\end{center} 
\end{figure}

For the numerical implementation, a simple three-band  effective mass model for
the electronic structure of silicon is used. The $X$ electrons are described by an
 multi-valley picture with identical values for transverse and
longitudinal effective mass, and the $\Gamma_{v}$ holes as well as the virtual
$\Gamma_{c}$ states used in the indirect transitions are modelled by single
effective mass bands. The holes are modelled by a single effective masses corresponding to
heavy and light holes. The band structure and interaction parameters used in the
numerical examples are listed in Tab. \ref{tab:band_parameters}. Fig. \ref{fig:indabs} displays the close agreement between NEGF and
FGR for equilibrium bulk absorption close to the indirect band edge and the characteristic fingerprints of the involved TO and LA phonon modes.  
 
\subsection{Thin film devices}
In the case of ultra-thin-film and especially hetero-multi-layer devices, translational invariance does no longer apply in 
growth direction. For such a system, the appropriate representation of the field operators has the form
\begin{align}
\hat{\Psi}_{b}(\mathbf{r},t)=\sum_{\mathbf{k}_{\parallel},i}\psi_{ib\mathbf{k}_{\parallel}}(\mathbf{r})
\hat{c}_{ib\mathbf{k}_{\parallel}}(t),\label{eq:thinfilmfieldop}
\end{align}
with the basis functions
\begin{align}
\psi_{in\mathbf{k}_{\parallel}}(\mathbf{r})=\varphi_{i\mathbf{k}_{\parallel}}(\mathbf{r})
u_{n\mathbf{k}_{0}}(\mathbf{r}),\label{eq:basis}
\end{align}
where $\varphi_{i\mathbf{k}_{\parallel}}$ is the envelope basis function for
discrete spatial (layer) index $i$ (longitudinal) and transverse momentum
$\mathbf{k}_{\parallel}$,   $u_{n\mathbf{k}_{0}}$ is the  Bloch function of
bulk band $n$, centered on $\mathbf{k}_{0}$. In the case of a system with
large transverse extension, the envelope basis function can be written as
\begin{align}
\varphi_{i\mathbf{k}_{\parallel}}(\mathbf{r})=\frac{e^{i\mathbf{k}_{\parallel}\mathbf{r}_{\parallel}}}{\sqrt{\mathcal{A}}}\chi_{i}(z),
\end{align}
where $\mathbf{r}_{\parallel}=(x,y)$, $\mathcal{A}$ is the cross sectional area and
$\chi_{i}$ is the longitudinal envelope basis function. For the latter, finite element shape functions are a popular
choice \cite{steiger:thesis, kubis:09}. Here, we will use a simple finite
difference basis equivalent to a separate single band tight-binding approach
for each band \cite{lake:97,henrickson:02,jin:06},  
\begin{align} 
\chi_{i}(z)=\left[\theta(z-z_{i})-\theta(z-z_{i+1})\right]/\sqrt{\Delta}.
\end{align}
In the above basis, the Green's functions and self-energies have the potentially non-local representation ($O=G,\Sigma$)
\begin{align}
O_{b,b'}(\mathbf{r},\mathbf{r'};E)=\sum_{\mathbf{k}_{\parallel}}\sum_{i,j}\psi_{ib\mathbf{k}_{\parallel}}(\mathbf{r})
O_{ib,jb'}(\mathbf{k}_{\parallel};E)\psi_{jb'\mathbf{k}_{\parallel}}^{*}(\mathbf{r'}),
\end{align}
where the contour-ordered steady-state Green's functions are now defined as
\begin{align}
O_{ib,jb'}(\mathbf{k}_{\parallel};t-t')=\frac{1}{i\hbar}\left\langle
\hat{T}_{\mathcal{C}}\left\{\hat{c}_{ib\mathbf{k}_{\parallel}}(t) \hat{c}_{jb'\mathbf{k}_{\parallel}}^{\dagger}(t')\right\}
\right\rangle.
\end{align}
\begin{figure}[t]
\begin{center}
\includegraphics[width=7cm]{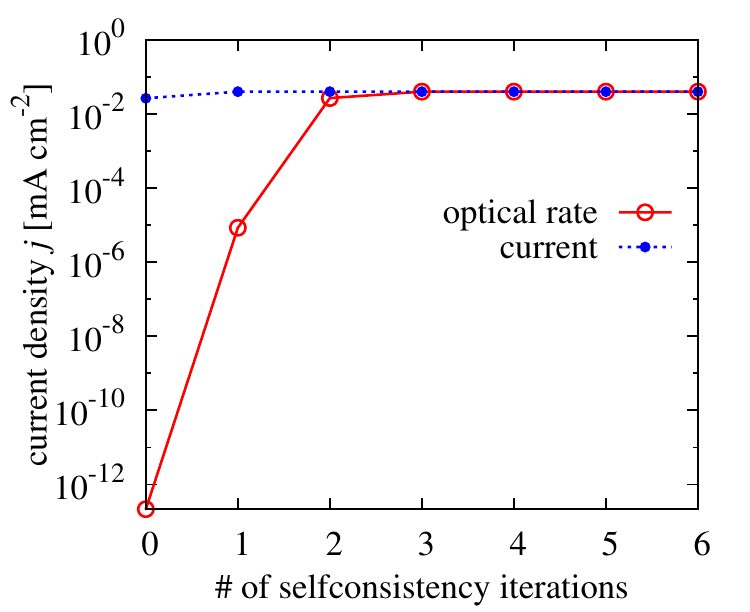} 
\caption{Convergence of the interband photogeneration rate (integrated over growth dimension) and of the resulting total interband current density in
the NEGF self-consistency iteration process (Fig. \ref{fig:selfconsit}).
\label{fig:convergence}}
\end{center}  
\end{figure} 
The conservation law corresponding to Eq. \eqref{eq:currcons} for the
divergence of the electron charge current between model layers $i-1$ and $i$
becomes 
\begin{align}
\frac{J_{i}-J_{i-1}}{\Delta}=&-\frac{2e}{\hbar\mathcal{A}\Delta}\sum_{{\mathbf k}}\int\frac{\d
E}{2\pi}\Big[\mathbf{\Sigma}^{R}\mathbf{G}^{<}\nonumber\\&-\mathbf{G}^{R}
 \mathbf{\Sigma}^{<}
 +\mathbf{\Sigma}^{<}\mathbf{G}^{A}-\mathbf{G}^{<}\mathbf{\Sigma}^{A}\Big]_{i,i}\label{eq:currdivlayer}\\
 &\equiv R_{rad}(z_{i}),
\end{align}
where $R_{rad}(z)$ is again the local radiative rate. Making use of the cyclic
property of the trace, the total radiative rate follows as 
\begin{align}
R_{rad}=&\frac{2}{\hbar\mathcal{A}}\int_{B}\frac{dE}{2\pi}\mathrm{Tr}\Big\{\sum_{\mathbf{k}}
\big[\mathbf{\Sigma}_{e\gamma}^{<}(\mathbf{k};E)\mathbf{G}^{>}(\mathbf{k};E)\nonumber\\&
-\mathbf{\Sigma}_{e\gamma}^{>}(\mathbf{k};E)\mathbf{G}^{<}(\mathbf{k};E)
\big]\Big\}\\
&\equiv \Delta\sum_{i}R_{rad}(z_{i})\label{eq:totratelayer}=\sum_{i}\left(J^{B}_{i}-J^{B}_{i-1}\right)\\
&\equiv J_{N}^{B}-J_{1}^{B} \equiv
\left\{\begin{array}{ll} J_{N},&B=CB,\\
J_{1},&B=VB.
\end{array}  \right.
\end{align}
 \begin{figure}[t!]
\begin{center}
\includegraphics[width=8cm]{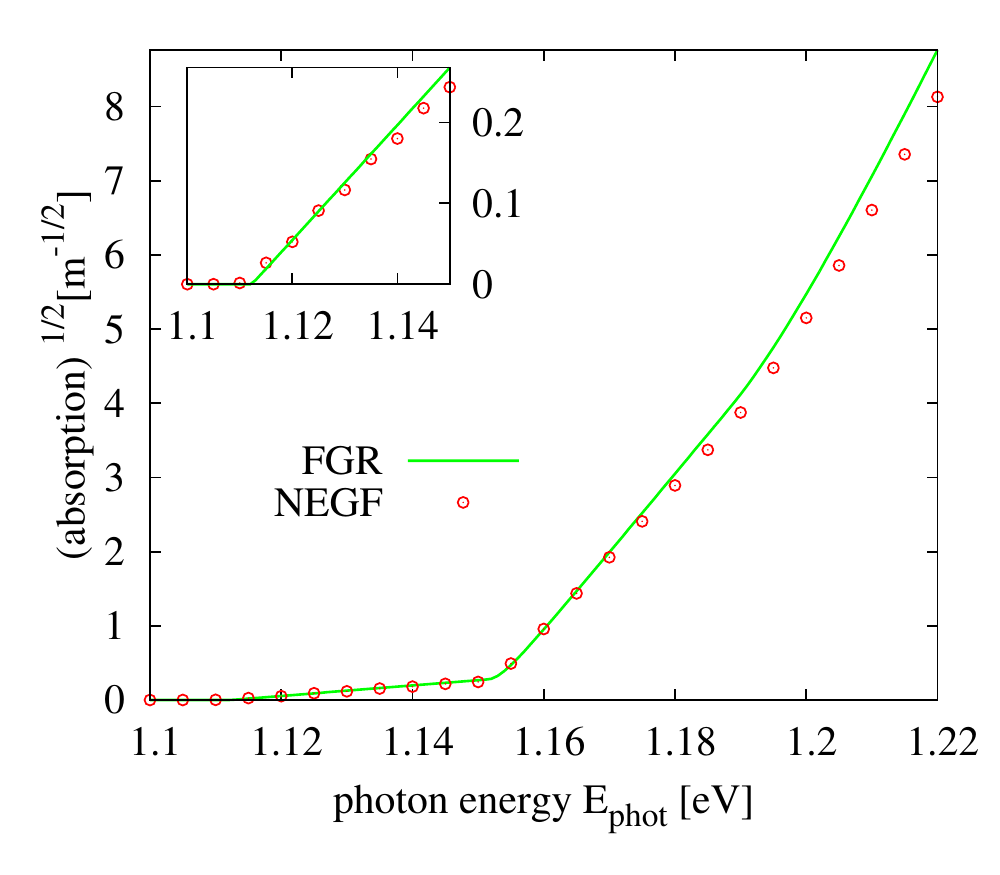}
\caption{Photon absorption rate for indirect optical transitions, computed with
the thin film model (dots) for a silicon $p$-$i$-$n$  diode with low doping
levels (10$^{16}$ cm$^{-3}$, in comparison with the Fermi Golden Rule rate for bulk
silicon (full line).
\label{fig:indabs_tf}}
\end{center} 
\end{figure}
In the case of the indirect gap material the total radiative rate is
\begin{align}
R_{e\gamma}
=&\frac{2}{\hbar\mathcal{A}}\int_{\Gamma_{c}}\frac{dE}{2\pi}\mathrm{Tr}\Big\{\sum_{\mathbf{k}_{\parallel}}
\big[\mathbf{\Sigma}_{e\gamma,\Gamma_{c}}^{<}(\mathbf{k}_{\parallel};E)\mathbf{G}_{\Gamma_{c}}^{>}(\mathbf{k}_{\parallel};E)
\nonumber\\
&-\mathbf{\Sigma}_{e\gamma,\Gamma_{c}}^{>}(\mathbf{k}_{\parallel};E)\mathbf{G}_{\Gamma_{c}}^{<}(\mathbf{k}_{\parallel};E)
\big]\Big\},\label{eq:totratelayer}
\end{align}
and the inter-valley phonon scattering rate reads
\begin{align} 
&R_{ep,\Gamma-X}
=\frac{2}{\hbar\mathcal{A}}\int_{\Gamma_{c}}\frac{dE}{2\pi}\mathrm{Tr}\Big\{\sum_{\mathbf{k}_{\parallel}}
\big[\mathbf{\Sigma}_{ep(\Gamma-X),\Gamma_{c}}^{<}(\mathbf{k}_{\parallel};E)\nonumber\\
&\times\mathbf{G}_{\Gamma_{c}}^{>}(\mathbf{k}_{\parallel};E)
-\mathbf{\Sigma}_{ep(\Gamma-X),\Gamma_{c}}^{>}(\mathbf{k}_{\parallel};E)\mathbf{G}_{\Gamma_{c}}^{<}(\mathbf{k}_{\parallel};E)
\big]\Big\}.
\end{align}
The convergence of scattering rate and interband current as a function of
self-consistency iteration steps is shown in  Fig. \ref{fig:convergence}. The
convergence is fast due to the low carrier density in the absorbing 
quasi-intrinsic region. 

\begin{figure*}[t]
\begin{center}
(a)\includegraphics[width=7cm]{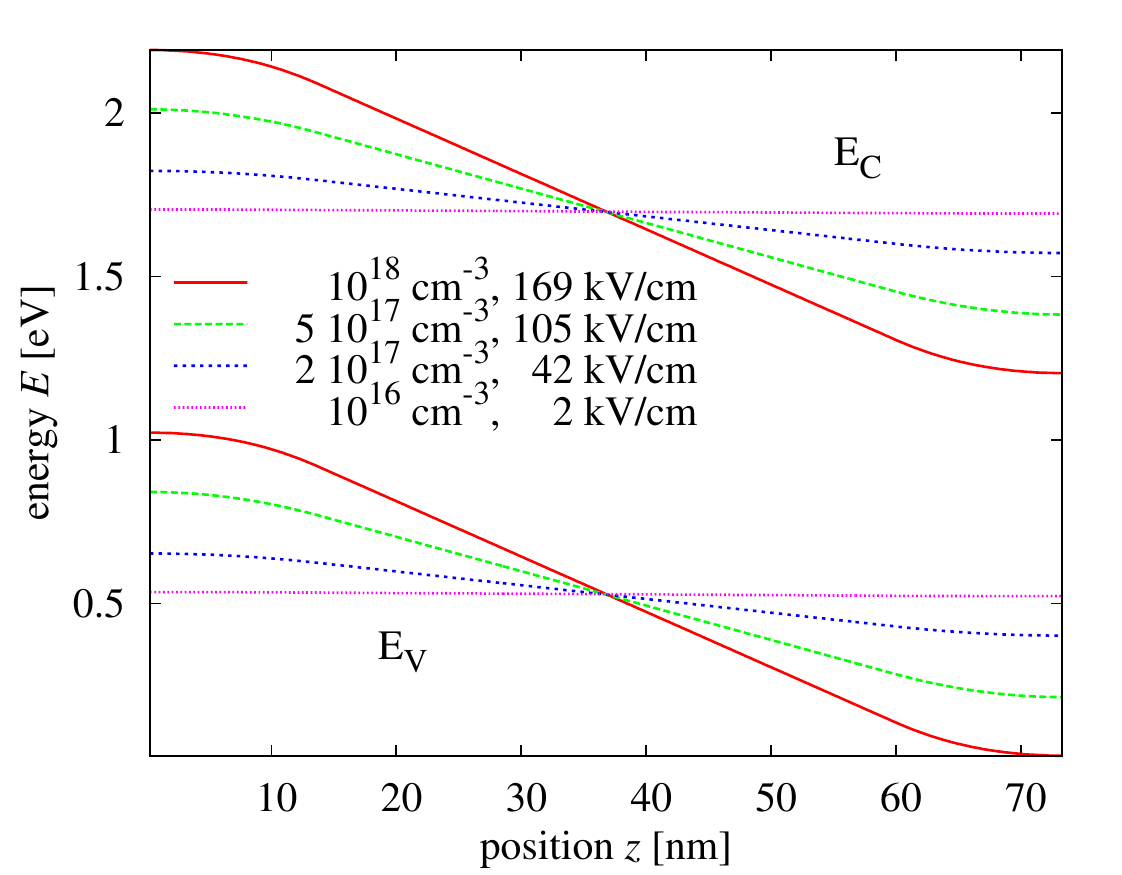}(b)\quad\includegraphics[width=6.2cm]{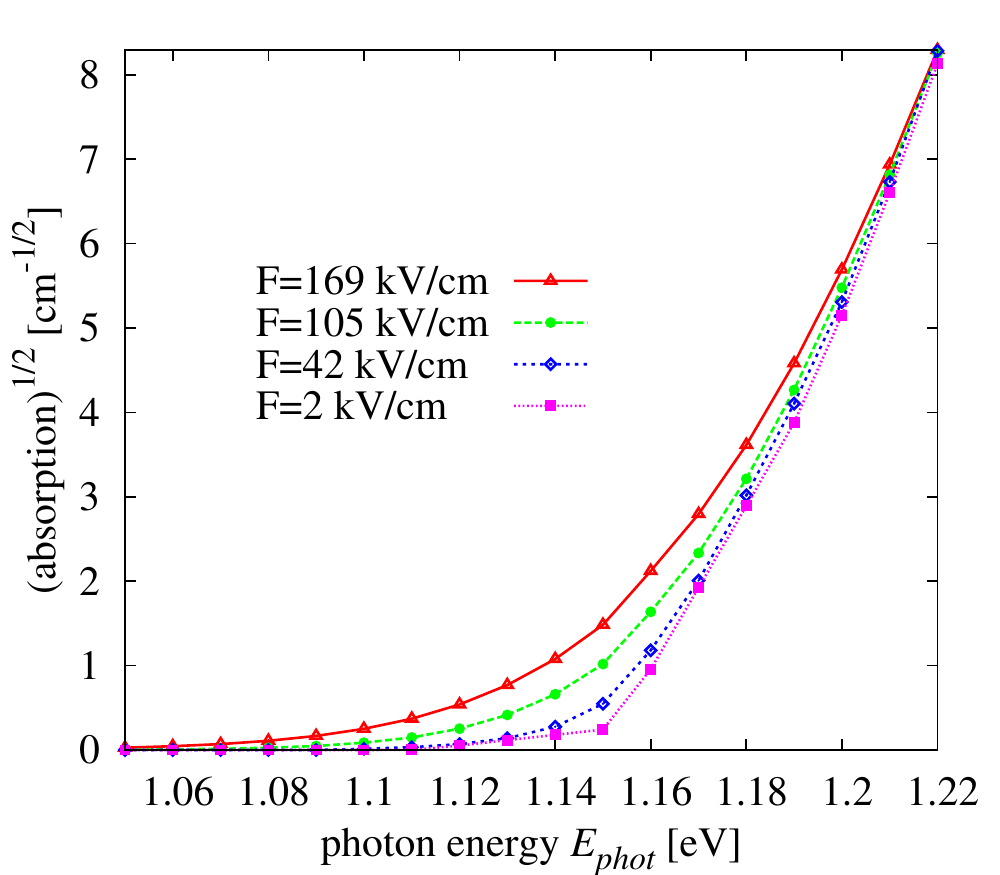}
\caption{(a) Band profile of a silicon $p$-$i$-$n$ junction from the
self-consistent solution of the NEGF-Poisson equations with the indicated doping
levels from $N_{d}=10^{16}$ cm$^{-3}$ to $N_{d}=10^{18}$ cm$^{-3}$. The corresponding
fields in the intrinsic region vary from $F=2$ kV/cm to  $F=169$ kV/cm. The
equilibrium Fermi level lies at 1.1 eV. (b) Near band gap optical absorption rate in the intrinsic region of a silicon $p$-$i$-$n$ junction, for various 
values of the built-in field corresponding to different doping levels.
\label{fig:field}}
\end{center}  
\end{figure*}

\begin{figure}[H] 
\begin{center}
(a)\includegraphics[width=4.4cm]{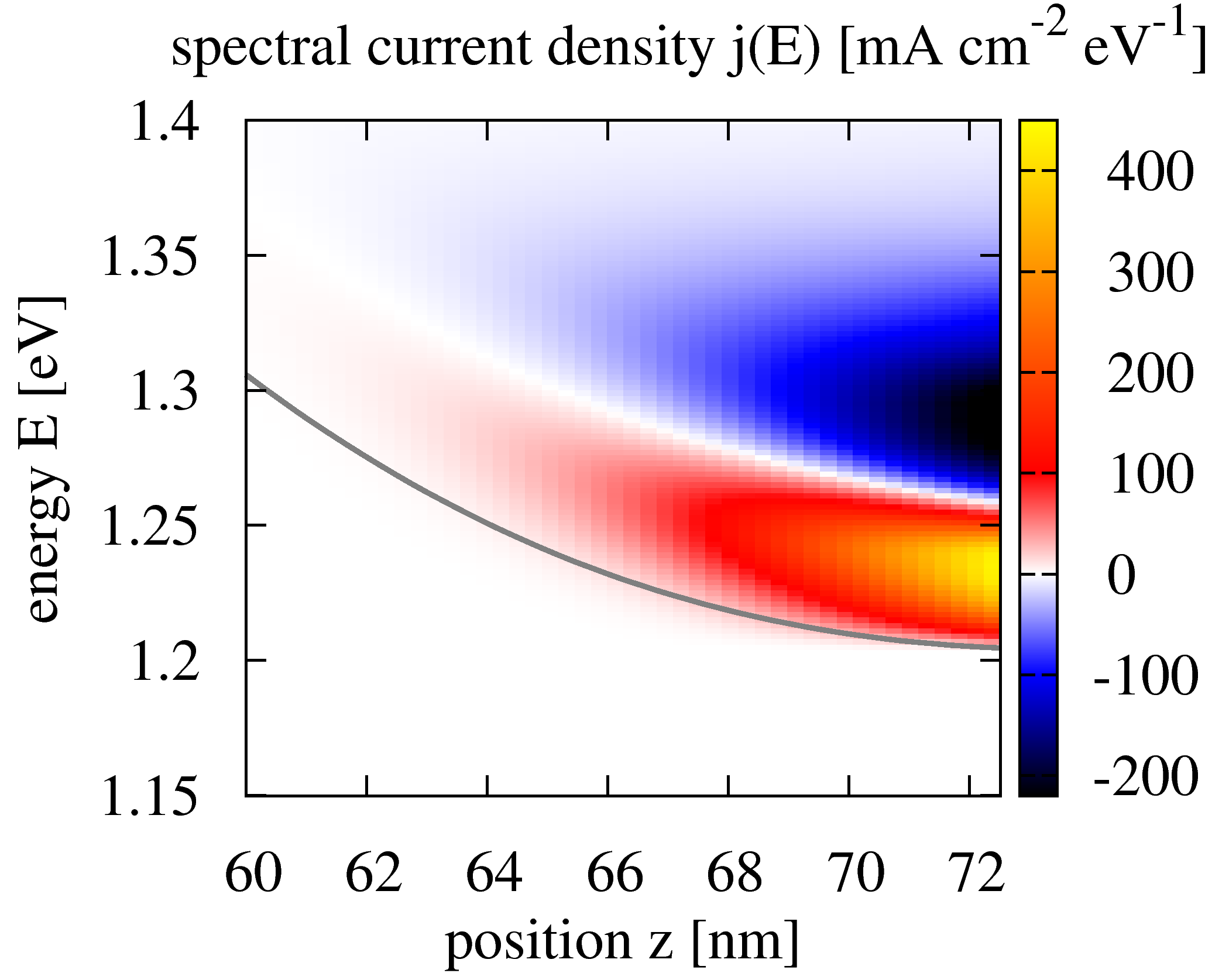}(b)\includegraphics[width=4.2cm]{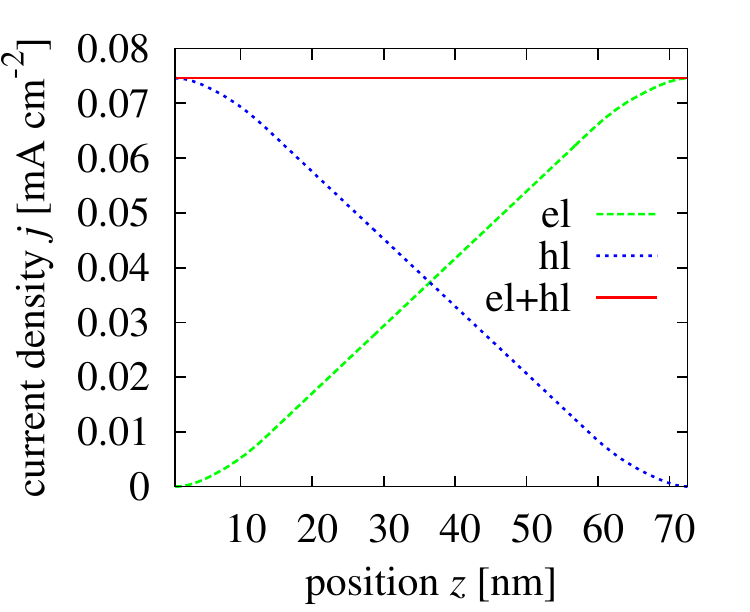}
\caption{(Color online) (a) Current spectrum of $X$-valley electrons close to
 the $n$-contact. (b) Integrated net current for electrons (el) and holes (hl) in the high-field device. The total current is perfectly conserved over
 the whole device.
\label{fig:local_current_int_nd24}}
\end{center}  
\end{figure}

\begin{figure*}[t!]
\begin{center}
(a)\includegraphics[width=8.2cm]{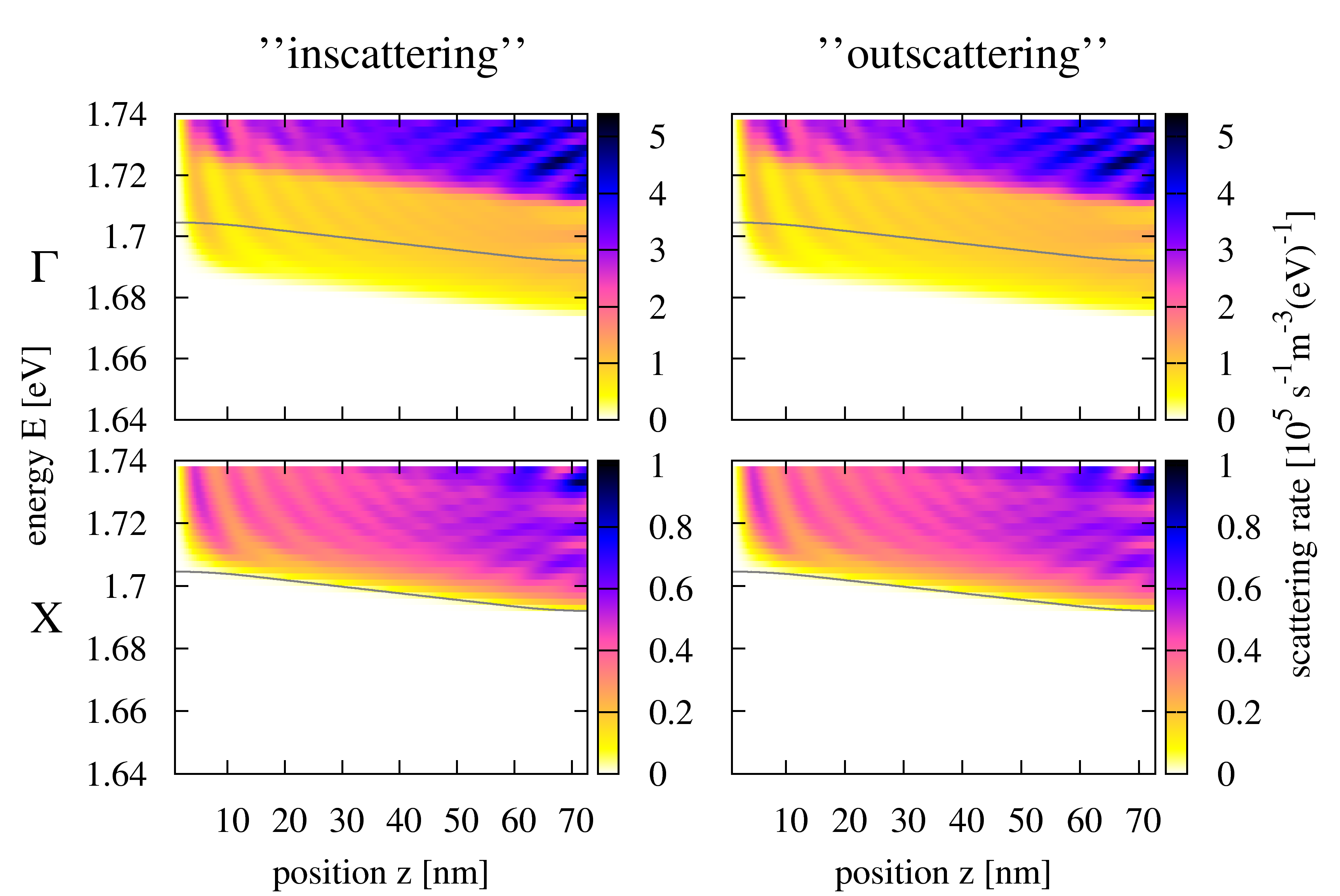}(b)\includegraphics[width=8.2cm]{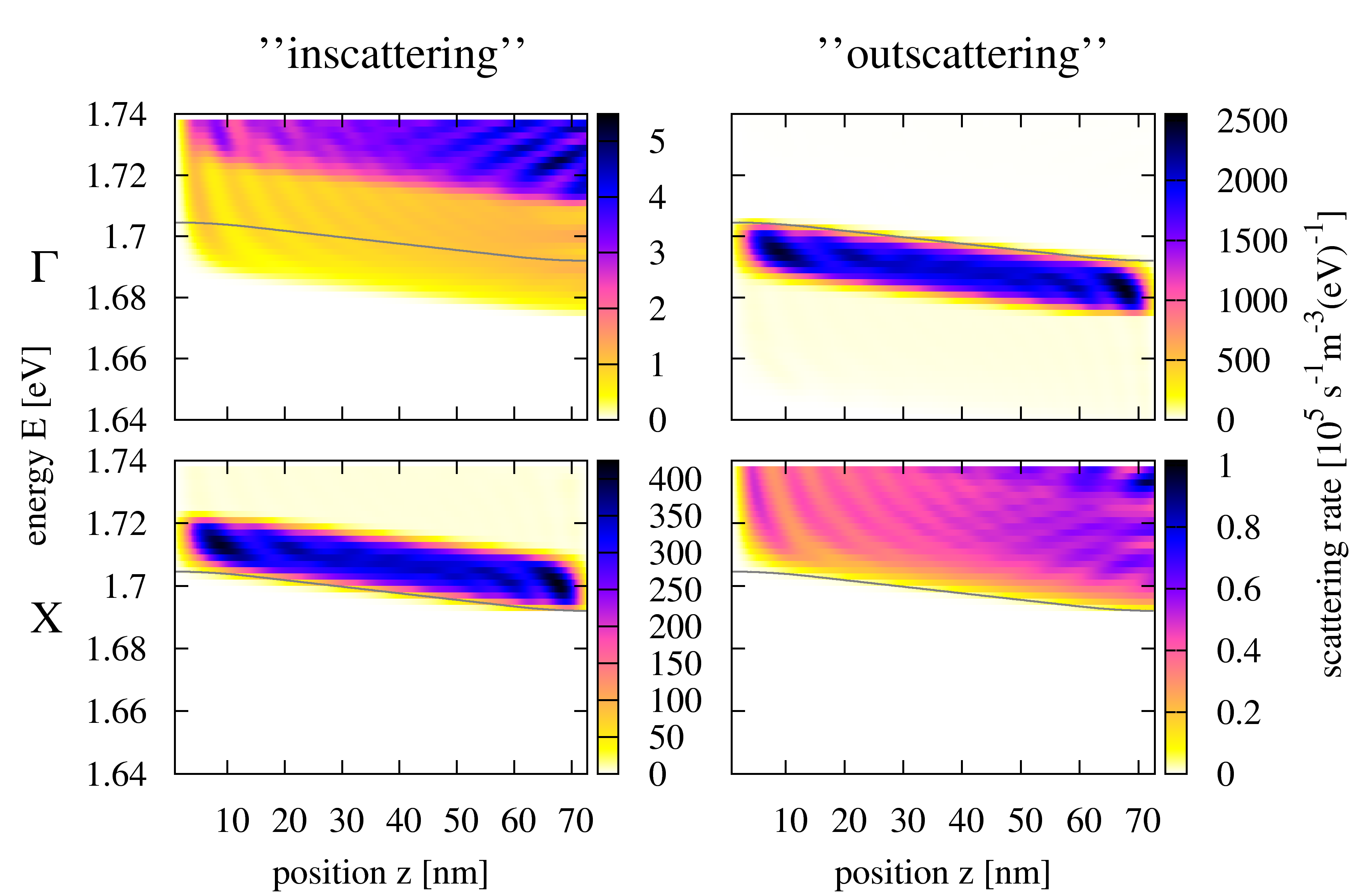}
\caption{(Color online) Phonon mediated electron transfer rate between virtual
($\Gamma_{c}$) and extended ($X$) conduction band states  (a) in the dark and
(b) under illumination with $E_{phot}=1.17$ eV, for the low-doping diode, where the intrinsic region extends
over the whole device. While in-and out-scattering are balanced in the dark,
there is a strong net charge transfer from $\Gamma_{c}$ to $X$ under illumination, and the inscattering rate reflects the spectral pattern of the
photogeneration.
\label{fig:phonscattrate_nd22}}
\end{center}  
\end{figure*}

Fig. \ref{fig:indabs_tf} shows the indirect optical
absorption for an quasi-intrinsic slab of the indirect gap semiconductor discussed above. In this case of near flat-band conditions, 
the bulk absorption is retrieved close to the band edge. For the more interesting case of a
ultra-thin bipolar junction with high doping,
the global absorption coefficient starts to deviate from the bulk value due to
the strong internal field. This so-called Franz-Keldysh effect is
displayed in Fig. \ref{fig:field}.
\begin{figure}[b]
\begin{center}
\includegraphics[width=8.2cm]{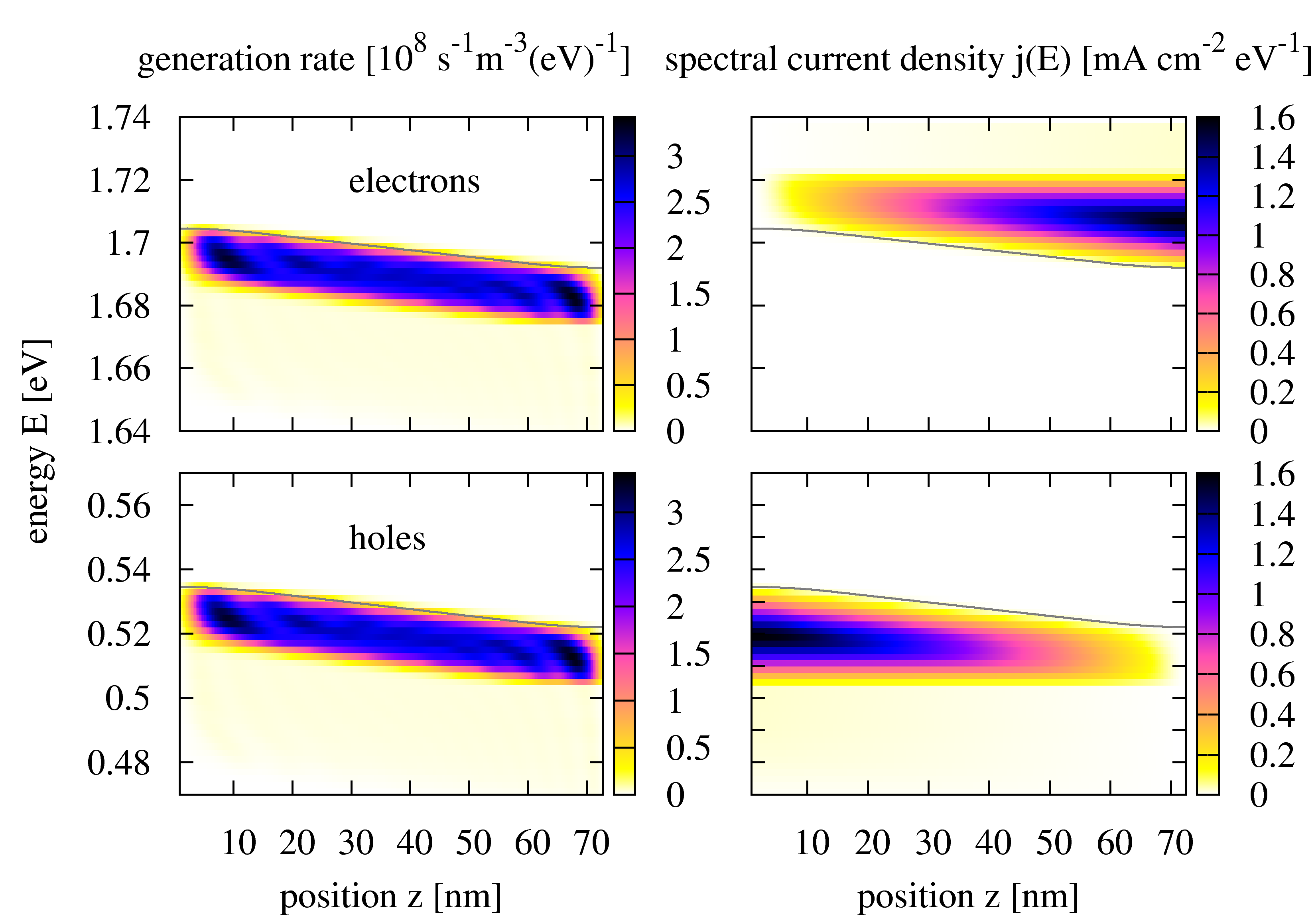}
\caption{(Color online) Energy resolved local generation rate in the virtual $\Gamma_c$-state and resulting local 
photocurrent spectrum in the $X$-valley band at very low internal field $F=2~kV/cm$. The grey lines indicate the band edges (X-valley for electrons).   
\label{fig:photscattrate_spectcurr_nd22}}
\end{center}  
\end{figure} 
\begin{figure}[t]
\begin{center}
\includegraphics[width=8.2cm]{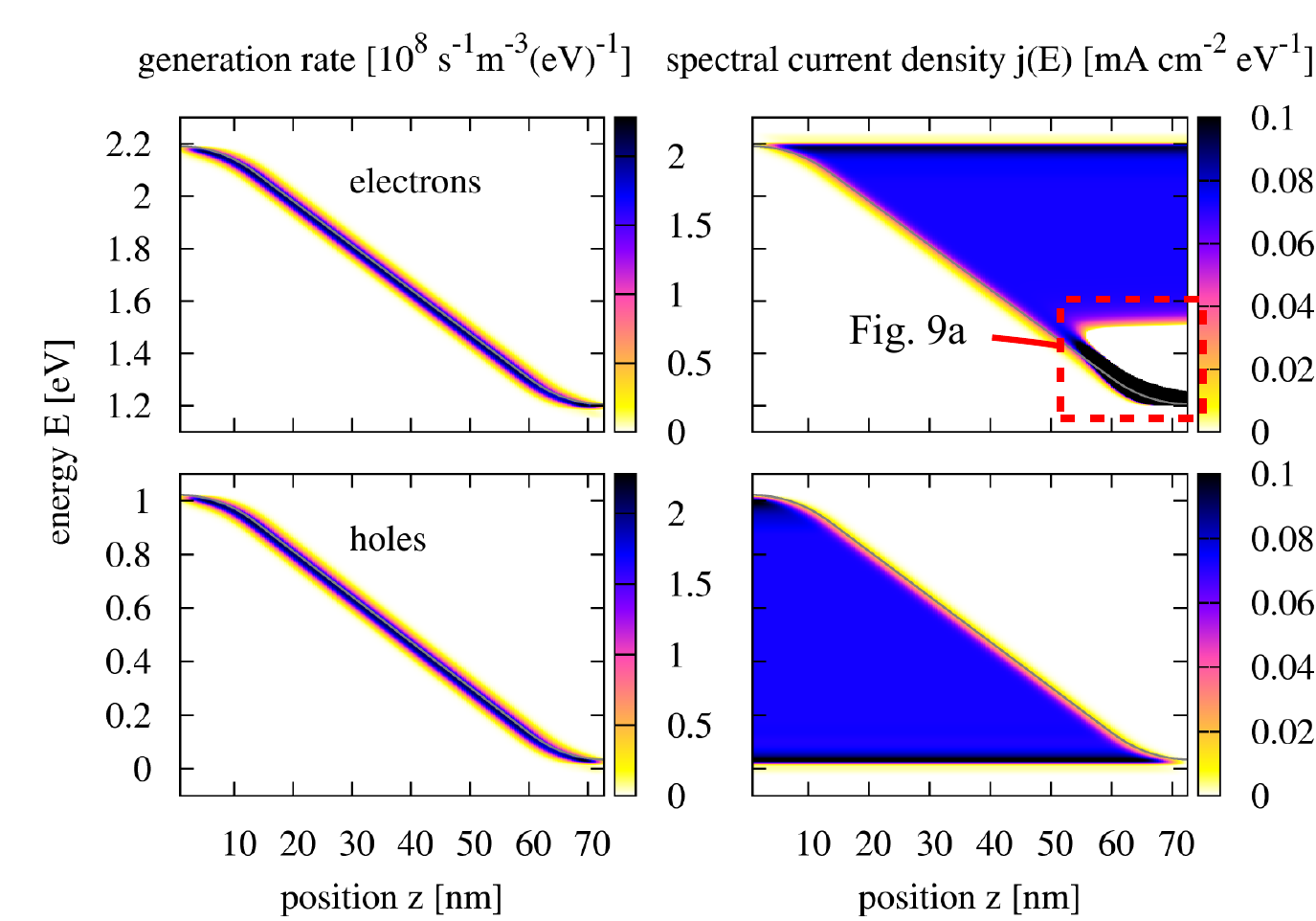}
\caption{(Color online) Same as Fig. \ref{fig:photscattrate_spectcurr_nd22}, but for a strong
internal field of $F=169~kV/cm$. The large deviation close to the $n$-contact is due to the electron-phonon intraband scattering current
 that vanishes upon energy integration over the band (white=negative, black=large positive, see Fig. \ref{fig:local_current_int_nd24}a).
\label{fig:photscattrate_spectcurr_nd24}}
\end{center}  
\end{figure}
Fig. \ref{fig:photscattrate_spectcurr_nd22} shows the energy resolved local generation rate in the 
virtual $\Gamma_c$-states and resulting local photocurrent spectrum in the $X$-valley band at a photon energy $E_{phot}=1.17$ eV and very low 
internal field $F=2$ kV/cm. The generation is relatively uniform throughout
the device, as well as the increase of electron and hole current components
towards the respective contacts. Fig. \ref{fig:photscattrate_spectcurr_nd24} shows the corresponding situation for the diode with strong
internal field $F=169$ kV/cm. In this case, the photogeneration and current
contributions are distributed over a large energy range, the latter due to the absence of efficient relaxation mechanisms that would confine current flow 
closer to the bandedge. In the vicinity of the $n$-contact, the high electron concentration results in large electron-phonon intraband scattering
current contributions (Fig. \ref{fig:local_current_int_nd24}a), which due to the opposite signs of in-
and outscattering components cancel upon energy integration over the band. The net current can thus be attributed exclusively to interband transitions, 
and the local sum of electron and hole contributions is perfectly conserved,
as shown in Fig. \ref{fig:local_current_int_nd24}b. 
 Since the photogeneration rate is intimately related to the $\Gamma-X$
 intervalley electron-phonon scattering rate via the self-consistent computation
 of Green's functions and self-energies, and the photocurrent results from
 photogenerated excess charge that is transferred via the scattering process
 with phonons to the extended $X$-valley states, the rates of this
 phonon-mediated charge transfer are expected to show the signature of the
 photogeneration process. Indeed, if the rate of electron scattering out of
 (into) the virtual $\Gamma_{c}$-states into (out of) the $X$-states is
 considered, there is a large net scattering from $\Gamma_{c}$ to $X$ under
 illumination, while in- and outscattering rates are balanced in the dark. This is shown in 
 Fig. \ref{fig:phonscattrate_nd22} for the case of the low-doping diode where the intrinsic region 
\begin{figure*}[t!]
\begin{center}
(a)\includegraphics[width=8.2cm]{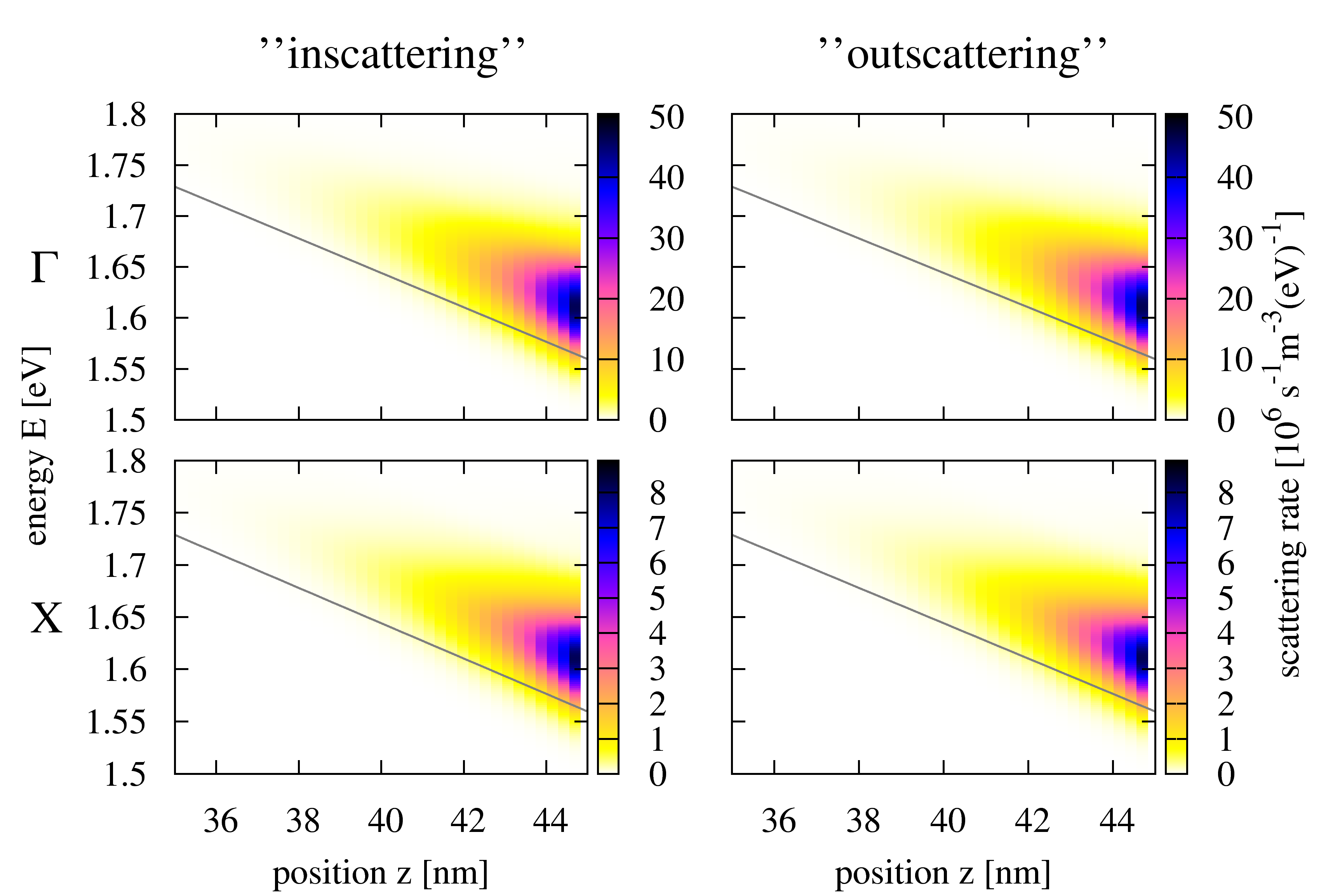}(b)\includegraphics[width=8.4cm]{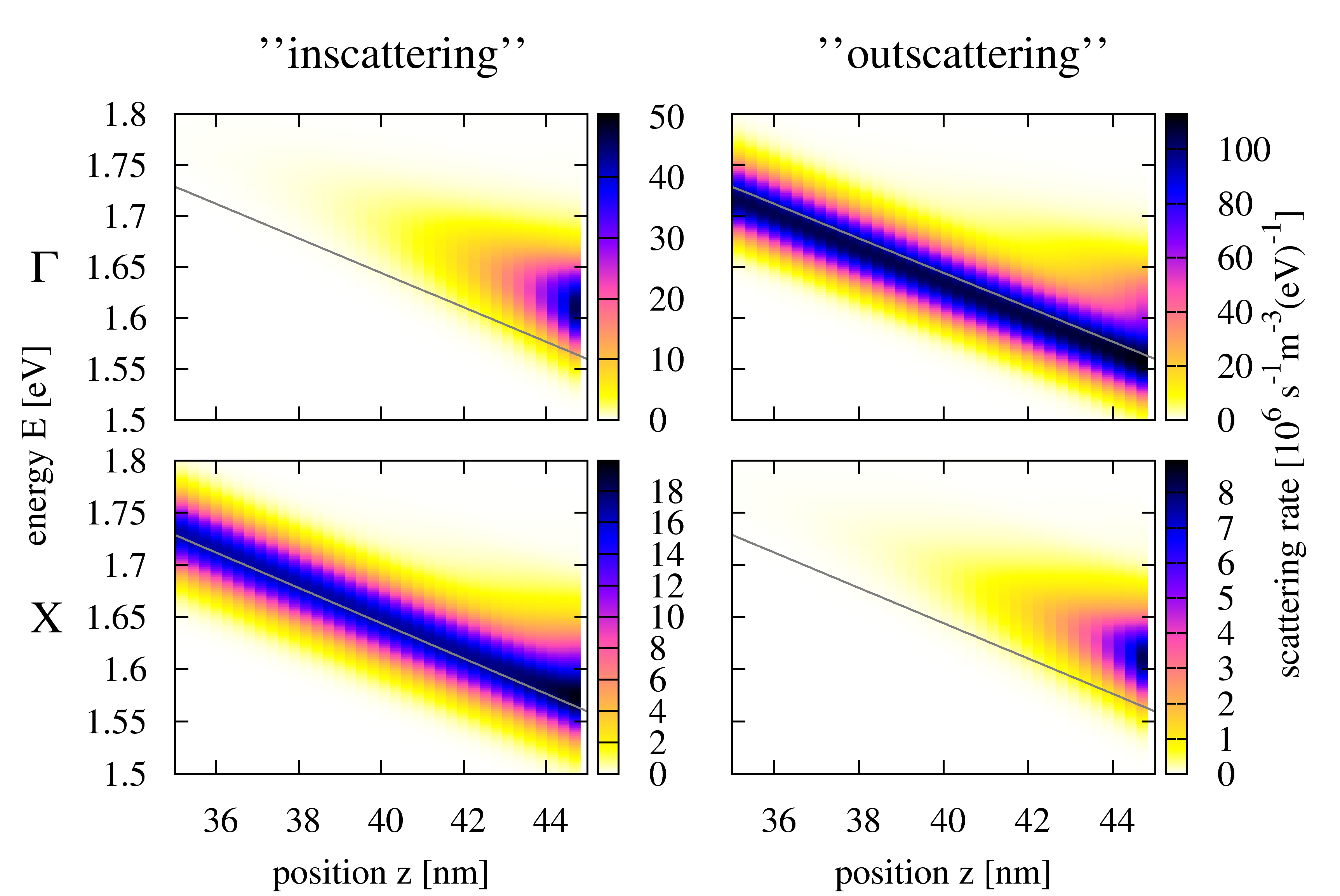}
\caption{(Color online) Phonon mediated electron transfer rate between virtual and extended
conduction band states (a) in the dark and (b) under illumination with
$E_{phot}=1.17$ eV in the center of the short intrinsic region of
the high-doping diode. Closer to the contacts, the scattering rate contributed by the photogeneration is masked by the large intraband relaxation
rate.
\label{fig:phonscattrate_nd24}}
\end{center}     
\end{figure*} 
 {\flushleft extends} over the entire device. In the case of strong doping, the electron-phonon scattering
 rate in the dark is much stronger close to the contacts, such that the effect
 becomes only visible in the central, intrinsic region of the junction. There, 
 the phonon-related charge transfer rate again reflects the local
 photogeneration rate, as displayed in Fig. \ref{fig:phonscattrate_nd24}.

\section{\label{sec:conclusion}Conclusions}
We have presented a novel theoretical approach to the description of
phonon-mediated photogeneration in indirect semiconductors. 
While compatible with the standard Fermi-Golden-Rule approach
in the quasi-equilibrium bulk limit, its range of validity extends to quantum transport in open systems involving arbitrary 
heterostructure states far from equilibrium and the effects of non-locality in the scattering processes, which
are important aspects of advanced photovoltaic and light emitting devices, where often spatial
and spectral resolution are required to gain access to a deeper understanding
of the device characteristics. Thanks to this versatility, the theoretical
framework lends itself to the modelling of indirect semiconductor based nanostructures with
potential applications in a multitude of optoelectronic devices, such as silicon-based quantum well solar cells.

\section*{Acknowledgements}
Financial support was provided by the German Federal Ministry
of Education and Research (BMBF) under Grant No. 03SF0352E.

 \appendix

 \section{\label{sec:appA}Interaction self-energies}

The  self-energies as defined by the Dyson equations
\eqref{eq:dyson1} and \eqref{eq:dyson2} encode the renormalization of the charge
carrier Green's functions due to the interactions with photons and phonons, i.e. generation, recombination and
 relaxation processes. The self-energies due to interactions can be evaluated
 from the perturbation expansion of the non-equilibrium statistical average defining the interacting
NEGF, using either Wick's theorem or Feynman diagrams. 
In the first case, the (contour ordered) self energy $\Sigma$ is derived from a
perturbation expansion of
the exponential in the definition of the contour ordered Green's function as the
nonequilibrium expectation value of single-carrier operators, 
\begin{align}
G_{\alpha,\alpha'}({\mathbf k;t,t'})&\equiv -\frac{i}{\hbar}\left\langle
\hat{T}_{C}\left\{e^{-\frac{i}{\hbar}\int_{C}ds H'(s)}\hat{c}_{\alpha,{\mathbf k
}}(t)\hat{c}^{\dagger}_{\alpha',{\mathbf k}}(t')\right\}\right\rangle,
\label{eq:selfenpert}
\end{align}
 $\alpha=n$ and $\mathbf{k}=(k_{x},k_{y},k_{z})$ (bulk) or $\alpha=i,n$
 and $\mathbf{k}=(k_{x},k_{y})\equiv\mathbf{k}_{\parallel}$ (thin film). In the
 following, the Hamiltonian terms for the perturbative interaction of electrons with photons and phonons
 shall be discussed and used in the derivation of the corresponding self-energy
 expressions for bulk and thin-film systems.
 
 \subsection{Interaction Hamiltonian}
 
 \paragraph{Electron-photon interaction}
 For the electron-photon interaction, the perturbation Hamiltonian is given via
 the linear coupling to the vector potential operator of the electromagnetic field  $\hat{{\mathbf A}}$,
\begin{align}
 \hat{H}_{e\gamma}=-\frac{e}{m_{0}}\hat{{\mathbf A}}\cdot\hat{{\mathbf
 p}}\label{eq:elphotham}
\end{align}
with $\hat{{\mathbf p}}$ the momentum operator and 
\begin{align}
\hat{\mathbf{A}}({\mathbf r},t)=&\sum_{\lambda,{\mathbf
q}}\left[\mathbf{A}_{0}(\lambda,\mathbf{q}) \hat{b}_{\lambda,{\mathbf
q}}(t)+\mathbf{A}_{0}^{*}(\lambda,-\mathbf{q}) 
\hat{b}_{\lambda,{-\mathbf q}}^{\dagger}(t)\right]\nonumber\\
&\times e^{i{\mathbf q}{\mathbf
r}},\label{eq:photfieldop}\\
\mathbf{A}_{0}(\lambda,\mathbf{q})=&\frac{\hbar}{\sqrt{2\epsilon_{0}V\hbar\omega_{\mathbf{q}}}}
\boldsymbol{\epsilon}_{\lambda{\mathbf q}},
\end{align}
where ${\mathbf \epsilon}_{\lambda{\mathbf q}} $ is the polarization of the
photon with wave vector 
${\mathbf q}$ and energy $\hbar\omega_{\mathbf{q}}$ added to or removed from photon
mode $(\lambda,\mathbf{q})$ by the bosonic creation and annihilation operators 
\begin{align}
\hat{b}_{\lambda,{\mathbf q}}^{\dagger}(t)=&\hat{b}_{\lambda,{\mathbf
q}}^{\dagger}e^{i\omega_{{\mathbf q}}t},\quad \hat{b}_{\lambda,{\mathbf
q}}(t)=\hat{b}_{\lambda,{\mathbf q}}e^{-i\omega_{{\mathbf q}}t},
\end{align}
and $V$ is the absorbing volume. 

For numerical evaluation, the Hamiltonian is reformulated in a suitable
representation using the field operators given in \eqref{eq:bulkfieldop} (bulk)
and \eqref{eq:thinfilmfieldop} (thin film), respectively,
 \begin{align}
\mathcal{H}_{e\gamma}(t)&=\int d^{3}r 
\hat{\Psi}^{\dagger}(\mathbf{r},t)\hat{H}_{e\gamma}\hat{\Psi}(\mathbf{r},t).
\end{align}
In the bulk case, the above expression results in\footnote{The photon momentum
is neglected as compared to the electron quasi-momentum.}
\begin{align}
\mathcal{H}_{e\gamma}(t)&=\sum_{\mathbf{q},\lambda}\sum_{n,m}\sum_{\mathbf{k},\mathbf{k}'}
\mathcal{M}^{e\gamma}_{n,m}(\mathbf{k},\mathbf{k}',\mathbf{q},\lambda)\nonumber\\
&\times\hat{c}^{\dagger}
_{n\mathbf{k}}(t)\hat{c}_{m\mathbf{k}'}(t)\left[\hat{b}_{\lambda,\mathbf{q}}
e^{-i\omega_{\mathbf{q}} t}+\hat{b}_{\lambda,-\mathbf{q}}^{\dagger}
e^{i\omega_{\mathbf{q}} t}\right],
\end{align}
where the matrix element for interband transitions ($n\neq m$) is obtained from
a $\mathbf{k}\cdot \mathbf{p}$ - type approximation
\cite{steiger:thesis}, 
\begin{align}
&\mathcal{M}^{e\gamma}_{n,m}(\mathbf{k},\mathbf{k}',\mathbf{q},\lambda)\equiv
-\frac{e}{m_{0}}A_{0}(\lambda,\mathbf{q})\nonumber\\
&\times\int d^3
 r\psi^{*}_{n\mathbf{k}}(\mathbf{r})\big(e^{i\mathbf{q}\mathbf{r}}
 \boldsymbol{\epsilon}_{\lambda{\mathbf q}}\cdot\hat{\mathbf{p}}\big)
 \psi_{m\mathbf{k}'}(\mathbf{r}) \\
  &\approx-\frac{e}{m_{0}}A_{0}(\lambda,\mathbf{q})
  \delta(\mathbf{k}'+\mathbf{q}-\mathbf{k})
  (\boldsymbol{\epsilon}_{\lambda{\mathbf q}}\cdot
 \mathbf{p}_{nm}),
 \label{eq:coupling_photons_bulk}
\end{align}
with the Bloch function momentum matrix element 
\begin{align}
\mathbf{p}_{nm}=\int_{\Omega}\frac
{d^3\tilde{r}}{\Omega}u_{n\mathbf{k}_{0}}(\tilde{\mathbf{r}})\hat{\mathbf{p}}u_{m\mathbf{k}_{0}'}(\tilde{\mathbf{r}}),
\end{align}
where $\Omega$ denotes the unit-cell volume. This gives the final bulk
expression
\begin{align}
\mathcal{H}_{e\gamma}(t)&=\sum_{\mathbf{q},\lambda}\sum_{n,m}\sum_{\mathbf{k}}
\mathcal{M}^{e\gamma}_{n,m}(\mathbf{k},\mathbf{q},\lambda)\nonumber\\
&\times\hat{c}^{\dagger}
_{n\mathbf{k}}(t)\hat{c}_{m\mathbf{k}}(t)\left[\hat{b}_{\lambda,\mathbf{q}}
e^{-i\omega_{\mathbf{q}} t}+\hat{b}_{\lambda,-\mathbf{q}}^{\dagger}
e^{i\omega_{\mathbf{q}} t}\right].
\end{align}

For devices with broken translational invariance in the transport dimension, the representation of the 
electron-photon Hamiltonian \eqref{eq:elphotham} in the real-space effective mass basis
\eqref{eq:basis} acquires the similar from
 \begin{align}
\mathcal{H}_{e\gamma}(t)&=\sum_{\mathbf{q},\lambda}\sum_{in,jm}\sum_{\mathbf{k}_{\parallel}}
\mathcal{M}^{e\gamma}_{in,jm}(\mathbf{k}_{\parallel},\mathbf{q},\lambda)\nonumber\\
&\times\hat{c}^{\dagger}
_{in\mathbf{k}_{\parallel}}(t)\hat{c}_{jm\mathbf{k}_{\parallel}}(t)\left[\hat{b}_{\lambda,\mathbf{q}}
e^{-i\omega_{\mathbf{q}} t}+\hat{b}_{\lambda,-\mathbf{q}}^{\dagger}
e^{i\omega_{\mathbf{q}} t}\right],
\end{align}
 where
 \begin{align}
\mathcal{M}^{e\gamma}_{in,jm}(\mathbf{k}_{\parallel},\mathbf{q},\lambda)
  &\approx-\frac{e}{m_{0}}A_{0}(\lambda,\mathbf{q})
 M_{ij}(q_{z})(\boldsymbol{\epsilon}_{\lambda{\mathbf q}}\cdot
 \mathbf{p}_{nm}),
 \label{eq:coupling_photons_tf}
\end{align}
 with
\begin{align}
M_{ij}(q_{z})=\int dz \chi^{*}_{i}(z)e^{i
q_{z}z}\chi_{j}(z)=e^{iq_{z}z_{i}}\delta_{ij}.
\end{align}

 \paragraph{Electron-phonon interaction}
 The vibrational degrees of freedom of the system are described in terms of the coupling of the force field 
of the electron-ion potential $V_{ei}$ to the 
quantized field $\boldsymbol{\mathcal{\hat{U}}}$ of the ionic displacement \cite{schaefer:02}, 

\begin{align}
\hat{H}_{ep}({\mathbf
r},t)=&\sum_{\mathbf{L},\boldsymbol{\kappa}}\boldsymbol{\mathcal{\hat{U}}}(\mathbf{L}+\boldsymbol{\kappa},t)\cdot
\nabla V_{ei}[\mathbf{r}-(\mathbf{L}+\boldsymbol{\kappa})],
\end{align}
with the displacement field given by the Fourier expansion
\begin{align}
\mathcal{\hat{U}}_{\alpha}(\mathbf{L}\boldsymbol{\kappa},t)=&\sum_{\Lambda,\mathbf{Q}}
\mathcal{U}_{\alpha\boldsymbol{\kappa}}(\Lambda,\mathbf{Q})
e^{i\mathbf{Q}\cdot(\mathbf{L}+\boldsymbol{\kappa})}
\big[\hat{a}_{\Lambda,\mathbf{Q}}(t)+\hat{a}^{\dagger}_{\Lambda,-\mathbf{Q}}(t)\big],
\nonumber\\&(\alpha=x,y,z),
\end{align}
where the ion equilibrium position is $\mathbf{L}+\boldsymbol{\kappa}$,
with $\mathbf{L}$ the lattice position and $\boldsymbol{\kappa}$ the relative
position of a specific basis atom at this lattice site, and
$\hat{a}_{\Lambda,{\mathbf Q}},\hat{a}_{\Lambda,{\mathbf Q}}^{\dagger}$ are 
the bosonic creation and annihilation operators for a (bulk) phonon mode with polarization
$\Lambda$ and wave vector ${\mathbf Q}$ in the first Brillouin zone. The potential felt by electrons 
in heterostructure states due to coupling to \emph{bulk} phonons can thus be
written as
\begin{equation}
  \hat{H}_{ep}({\mathbf r},t)=\frac{1}{\sqrt{V}}\sum_{\Lambda{\mathbf
  Q}}U_{\Lambda,{\mathbf Q}}e^{i{\mathbf Q}\cdot{\mathbf
  r}}\{\hat{a}_{\Lambda,{\mathbf Q}}(t)+ \hat{a}_{\Lambda,-{\mathbf
  Q}}^{\dagger}(t)\},
\label{eq:el_phonint}
\end{equation}
where $\mathbf{r}$ is the electron coordinate, and
$U_{\Lambda,\mathbf{Q}}$ are related to the Fourier coefficients of
the electron-ion potential \cite{mahan:90}.

The effective-mass Hamiltonian for electron-phonon interaction
 is obtained from \eqref{eq:el_phonint} in analogy to the electron-photon interaction,
 with the bulk result
 \begin{align}
  \mathcal{H}_{ep}(t)=&
  \sum_{{\mathbf
  Q},\Lambda}\sum_{n,{\mathbf
  k}}\mathcal{M}^{ep}
  (\mathbf{Q},\Lambda) \hat{c}_{n{\mathbf
  k}}^{\dagger}(t)\hat{c}_{n\mathbf{k}
 -\mathbf{Q}}(t)\nonumber\\ &\times\big[\hat{a}
  _{\Lambda,{\mathbf Q}}e^{-i\Omega_{\Lambda,\mathbf{Q}}t}
  +\hat{a}_{\Lambda,-{\mathbf
  Q}}^{\dagger}e^{i\Omega_{\Lambda,\mathbf{Q}}t}\big],
\end{align}
with
 \begin{align}
 \mathcal{M}^{ep}(\mathbf{Q},\Lambda)&=\frac{U_{\Lambda,{\mathbf
 Q}}}{\sqrt{V}}\label{eq:ep_coupling_bulk},
 \end{align}
 and the thin-film expression
\begin{align}
  \mathcal{H}_{ep}(t)=&
  \sum_{{\mathbf
  Q},\Lambda}\sum_{n}\sum_{{\mathbf
  k}_{\parallel},i}\mathcal{M}_{i}^{ep}
  (\mathbf{Q},\Lambda) \hat{c}_{in{\mathbf
  k}_{\parallel}}^{\dagger}(t)\hat{c}_{in\mathbf{k}
  _{\parallel}-\mathbf{Q}_{\parallel}}(t)\nonumber\\ &\times\big[\hat{a}
  _{\Lambda,{\mathbf Q}}e^{-i\Omega_{\Lambda,\mathbf{Q}}t}
  +\hat{a}_{\Lambda,-{\mathbf
  Q}}^{\dagger}e^{i\Omega_{\Lambda,\mathbf{Q}}t}\big],
\end{align}
 where
 \begin{align}
 \mathcal{M}_{i}^{ep}(\mathbf{Q},\Lambda)&=\frac{U_{\Lambda,{\mathbf
 Q}}}{\sqrt{V}}e^{iQ_{z}z_{i}}.\label{eq:ep_coupling_tf}
 \end{align}
 For the $\Gamma-X$ intervalley scattering considered in the present discussion, the coupling reads
\begin{equation}
  |U_{\Lambda,\mathbf{Q}}|^{2}=\frac{\hbar (D_{iv}K)^{2}_{\sigma}}{2\rho
  \Omega_{\sigma}},
\label{eq:ivphoncoupling}
\end{equation}
where $\sigma$ labels the phonon mode, $D_{iv}$ is the associated
deformation potential and $K$ denotes the momentum transfer required for the
scattering between two valleys. 

 \subsection{\label{sec:app_se}Self-energy}
 
At this stage, any renormalizing effect of
the electronic system on the photons and phonons is neglected, i.e. the coupling to the bosons
corresponds to the connection to corresponding equilibrium reservoirs. While
this treatment is generally a good approximation in the case of phonons, it
is valid for the coupling to the photonic systems only in the case of low
absorption, i.e. weak coupling or very short absorber length. 
The equilibrium propagators for non-interacting photons and phonons in isotropic
media have the common form ($\alpha=\gamma,p$)
\begin{align}
  D_{\lambda}^{\alpha,\lessgtr}({\mathbf q};E) =&-2\pi
  i\Big[N^{\alpha}_{\lambda,{\mathbf q}}\delta(E\mp \hbar\omega_{{\mathbf q}})\nonumber\\
  &+(N^{\alpha}_{\lambda,{\mathbf
        q}}+1)\delta(E\pm\hbar\omega_{{\mathbf q}})\Big],\\
  D_{\lambda}^{\alpha,R/A}({\mathbf q};E)&=\frac{1}{E-\hbar\omega_{{\mathbf
  q}}\pm i\eta}-\frac{1}{E+\hbar\omega_{{\mathbf q}}\pm i\eta}.
\label{eq:equilphonen}
\end{align}
In the above expressions, $N^{\alpha}_{\lambda,\mathbf{q}}$ denotes the occupation of the respective
equilibrium boson modes, with the phonon occupation given by the Bose-Einstein
distribution at lattice temperature $T$,
\begin{align}
N^{p}_{\Lambda,\mathbf{Q}}=\left(e^{\beta\hbar\Omega_{\Lambda,\mathbf{Q}}}-1\right)^{-1},\quad\beta=(k_{B}T)^{-1},
\end{align}
and the photon occupation $N^{\gamma}_{\lambda,\mathbf{q}}$ introduced in Sec.
\ref{sec:dirintrans}. Inserting these propagators the Fock-term of the a general electron-boson
self-energy in the first self-consistent Born approximation provides the
steady-state components ($\alpha=\gamma,p$)

\begin{align}
\boldsymbol{\Sigma}_{e\alpha}^{\lessgtr}({\mathbf k};E)=&
\sum_{\lambda,\mathbf{q}}\boldsymbol{\mathcal{M}}^{e\alpha}({\mathbf
k},\mathbf{q},\lambda)
\big[N^{\alpha}_{\lambda,\mathbf{q}}\mathbf{G}^{\lessgtr}({\mathbf
k};E\mp\hbar\omega_{\lambda,\mathbf{q}})\nonumber\\
&+(N^{\alpha}_{\lambda,\mathbf{q}}
+1)\mathbf{G}^{\lessgtr} ({\mathbf
k}_{\parallel};E\pm\hbar\omega_{\lambda,\mathbf{q}})\big]\nonumber\\
&\times\boldsymbol{\mathcal{M}}^{e\alpha}({\mathbf
k}_{\parallel},-\mathbf{q},\lambda)\label{eq:intse_lg}
\end{align}
and 
\begin{align}
\mathbf{\Sigma}_{e\alpha}^{R,A}({\mathbf
k};E)&=i\int\frac{dE'}{2\pi}\frac{\boldsymbol{\Sigma}_{e\alpha}^{>}(\mathbf{k};E')
-\boldsymbol{\Sigma}_{e\alpha}^{<}(\mathbf{k}_{\parallel};E')}{E'-E\pm
i\eta}
\label{eq:intse_ra}
\end{align}
with the same definitions of indices and momentum as in \eqref{eq:selfenpert}.
Using a constant mode specific coupling strength, the intervalley bulk
(thin-film) electron-phonon scattering self-energy is further simplified to the
diagonal form used in the simulations,
\begin{align}
  \Sigma_{(ij,)b}^{\lessgtr}(E)=&\sum_{\sigma}\sum_{b'\neq b}\frac{\hbar
  (D_{iv}K)^{2}_{\sigma}}{2\rho
  \Omega_{\sigma}\Delta}f_{b'}(\delta_{i,j})\nonumber\\&\times\int
  \frac{d\mathbf{k}}{4\pi^{2}}\Big[N^{p}_{\sigma}
  G^{\lessgtr}_{(ij,)b'}(\mathbf{k};E\pm\hbar\Omega_{\sigma})\nonumber\\
  &+(N^{p}_{\sigma}+1)
  G^{\lessgtr}_{(ij,)b'}(\mathbf{k};E\mp\hbar\Omega_{\sigma})\Big],\label{eq:phonintse}
\end{align}
 where $b,b'\in \left \{\Gamma_{c},X\right\}$.
 
\vspace{3cm}
 \bibliographystyle{apsrev4-1}
\bibliography{/home/aeberurs/Biblio/bib_files/negf,/home/aeberurs/Biblio/bib_files/aeberurs,/home/aeberurs/Biblio/bib_files/generation,/home/aeberurs/Biblio/bib_files/scqmoptics,/home/aeberurs/Biblio/bib_files/pv,/home/aeberurs/Biblio/bib_files/qd}
 
\end{document}